\documentstyle[sprocl]{article}
\input epsf

\global\arraycolsep=2pt

\begin{document}

\begin{titlepage}

\begin{flushright}
CERN-TH/96-281\\
hep-ph/9610266
\end{flushright}

\vspace{2cm}

\begin{center}
\Large\bf Heavy-Quark Effective Theory
\end{center}

\vspace{2cm}

\begin{center}
Matthias Neubert\\
{\sl Theory Division, CERN, CH-1211 Geneva 23, Switzerland}
\end{center}

\vspace{1cm}

\begin{abstract}
We give an introduction to the heavy-quark effective theory and the
$1/m_Q$ expansion, which provide the modern framework for a
systematic, model-independent description of the properties and
decays of hadrons containing a heavy quark. We discuss the
applications of these concepts to spectroscopy and to the weak
decays of $B$ mesons.
\end{abstract}

\vspace{1.5cm}

\begin{center}
Lectures presented at the\\
34th International School of Subnuclear Physics\\
{\it Effectives Theories and Fundamental Interactions}\\
Erice, Italy, 3-12 July 1996
\end{center}

\vspace{2cm}

\noindent
CERN-TH/96-281\\
October 1996
\vfil

\end{titlepage}

\thispagestyle{empty}
\vbox{}
\newpage

\setcounter{page}{1}


\title{HEAVY-QUARK EFFECTIVE THEORY}

\author{MATTHIAS NEUBERT}

\address{Theory Division, CERN, CH-1211 Geneva 23, Switzerland}

\maketitle\abstracts{
We give an introduction to the heavy-quark effective theory and the
$1/m_Q$ expansion, which provide the modern framework for a
systematic, model-independent description of the properties and
decays of hadrons containing a heavy quark. We discuss the
applications of these concepts to spectroscopy and to the weak
decays of $B$ mesons.}

\section{Introduction}
 
The weak decays of hadrons containing a heavy quark are employed for
tests of the Standard Model and measurements of its parameters. They
offer the most direct way to determine the weak mixing angles, to
test the unitarity of the Cabibbo--Kobayashi--Maskawa (CKM) matrix,
and to explore the physics of CP violation. At the same time,
hadronic weak decays also serve as a probe of that part of
strong-interaction phenomenology which is least understood: the
confinement of quarks and gluons inside hadrons.

\begin{figure}[htb]
   \epsfxsize=5cm
   \centerline{\epsffile{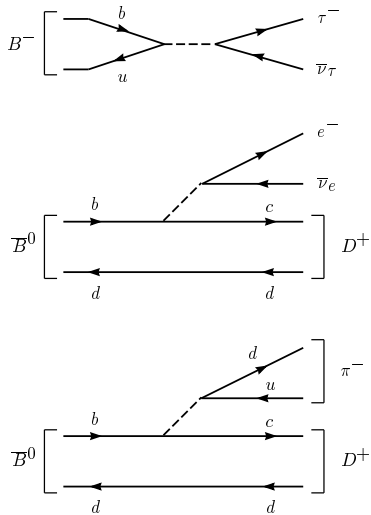}}
\caption{\label{fig:classes}
Examples of leptonic ($B^-\to\tau^-\bar\nu_\tau$), semileptonic
($\bar B^0\to D^+ e^-\bar\nu_e$), and non-leptonic ($\bar B^0\to
D^+\pi^-$) decays of $B$ mesons.}
\end{figure}

The structure of weak interactions in the Standard Model is rather
simple. Flavour-changing decays are mediated by the coupling of the
charged current to the $W$-boson field. At low energies, the
charged-current interaction gives rise to local four-fermion
couplings, whose strength is governed by the Fermi constant
\begin{equation}
   G_F = {g^2\over 4\sqrt{2} M_W^2} = 1.16639(2)~\mbox{GeV}^{-2} \,.
\end{equation}
According to the structure of the these interactions, the weak decays
of hadrons can be divided into three classes: leptonic decays, in
which the quarks of the decaying hadron annihilate each other and
only leptons appear in the final state; semileptonic decays, in which
both leptons and hadrons appear in the final state; and non-leptonic
decays, in which the final state consists of hadrons only.
Representative examples of these three types of decays are shown in
Fig.~\ref{fig:classes}.

\begin{figure}[htb]
   \epsfxsize=8.5cm
   \centerline{\epsffile{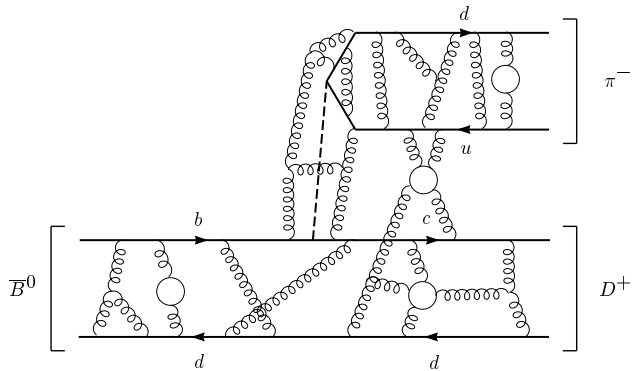}}
\caption{\label{fig:nonlep}
More realistic representation of a non-leptonic decay.}
\end{figure}

The simple quark-line graphs shown in this figure are a gross
oversimplification, however. In the real world, quarks are confined
inside hadrons, bound by the exchange of soft gluons. The simplicity
of the weak interactions is overshadowed by the complexity of the
strong interactions. A complicated interplay between the weak and
strong forces characterizes the phenomenology of hadronic weak
decays. As an example, a more realistic picture of a non-leptonic
decay is shown in Fig.~\ref{fig:nonlep}. Clearly, the complexity of
strong-interaction effects increases with the number of quarks
appearing in the final state. Bound-state effects in leptonic decays
can be lumped into a single parameter (a ``decay constant''), while
those in semileptonic decays are described by invariant form factors,
depending on the momentum transfer $q^2$ between the hadrons.
Approximate symmetries of the strong interactions help to constrain
the properties of these form factors. For non-leptonic decays, on the
other hand, we are still far from having a quantitative understanding
of strong-interaction effects even in the simplest decay modes.

Over the last decade, a lot of information on heavy-quark decays has
been collected in experiments at $e^+ e^-$ and hadron colliders. This
has led to a rather detailed knowledge of the flavour sector of the
Standard Model and many of the parameters associated with it. There
have been several great discoveries in this field, such as
$B^0$--$\bar B^0$ mixing~\cite{BBbar1,BBbar2}, charmless $B$
decays~\cite{btou1}$^-$\cite{Bpirho}, and rare decays induced by
penguin operators~\cite{BKstar,btos}. The experimental progress in
heavy-flavour physics has been accompanied by a significant progress
in theory, which was related to the discovery of heavy-quark symmetry
and the development of the heavy-quark effective theory (HQET). The
excitement about these developments is caused by the fact that they
allow (some) model-independent predictions in an area in which
``progress'' in theory often meant nothing more than the construction
of a new model, which could be used to estimate some
strong-interaction hadronic matrix elements. In these notes, we
explain the physical picture behind heavy-quark symmetry and discuss
the construction, as well as simple applications, of the heavy-quark
expansion. Because of lack of time, we will have to focus on some
particularly important aspects, emphasizing the main ideas and
concepts of the HQET. A more complete discussion of the applications
of this formalism to heavy-flavour phenomenology can be found in some
recent review articles~\cite{review,Shrev}. The reader is also
encouraged to consult the earlier review
papers~\cite{GeRev}$^-$\cite{Grorev} on the subject.

Hadronic bound states of a heavy quark with light constituents
(quarks, antiquarks and gluons) are characterized by a large
separation of mass scales: the heavy-quark mass $m_Q$ is much larger
than the mass scale $\Lambda_{\rm QCD}$ associated with the light
degrees of freedom. Equivalently, the Compton wave length of the
heavy quark ($\lambda_Q\sim 1/m_Q$) is much smaller than the size of
the hadron containing the heavy quark ($R_{\rm had}\sim
1/\Lambda_{\rm QCD}$). Our goal will be to separate the physics
associated with these two scales, in such a way that all dependence
on the heavy-quark mass becomes explicit. The framework in which to
perform this separation is the operator product expansion
(OPE)~\cite{Wils,Zimm}. The HQET provides us with a convenient
technical tool to construct the OPE. Before we start to explore in
detail the details of this effective theory, however, we should
mention two important reasons why it is desirable to separate short-
and long-distance physics in the first place:
\begin{itemize}
\item
A technical reason is that after the separation of short- and
long-distance phenomena we can actually calculate a big portion of
the relevant physics (i.e.\ all short-distance effects) using
perturbation theory and renorma\-li\-za\-tion-group techniques. In
particular, in this way we will be able to control all logarithmic
dependence on the heavy-quark mass.
\item
An important physical reason is that, after the short-distance
physics has been separated, it may happen that the long-distance
physics simplifies due to the realization of approximate symmetries,
which imply non-trivial relations between observables.
\end{itemize}
The second point is particularly exciting, since it allows us to make
statements beyond the range of applicability of perturbation theory.
Notice that here we are not talking about symmetries of the full QCD
Lagrangian, such as its local gauge symmetry, but approximate
symmetries realized in a particular kinematic situation. In
particular, we will find that an approximate spin--flavour symmetry
is realized in systems in which a single heavy quark interacts with
light degrees of freedom by the exchange of soft gluons. 

At this point it is instructive to recall a more familiar example of
how approximate symmetries relate the long-distance physics of
several observables. The strong interactions of pions are severely
constrained by the approximate chiral symmetry of QCD. In a certain
kinematic regime, where the momenta of the pions are much less than
1~GeV (the scale of chiral-symmetry breaking), the long-distance
physics of scattering amplitudes is encoded in a few ``reduced matrix
elements'', such as the pion decay constant. An effective low-energy
theory called chiral perturbation theory provides a systematic
expansion of scattering amplitudes in powers of the pion momenta, and
thus helps to derive the relations between different scattering
amplitudes imposed by chiral symmetry~\cite{Leut}. We will find that
a similar situation holds for the case of heavy quarks. Heavy-quark
symmetry implies that, in the limit where $m_Q\gg\Lambda_{\rm QCD}$,
the long-distance physics of several observables is encoded in few
hadronic parameters, which can be defined in terms of operator matrix
elements in the HQET.

\section{Heavy-Quark Symmetry}
\label{sec:2}

\subsection{The Physical Picture}
 
There are several reasons why the strong interactions of systems
containing heavy quarks are easier to understand than those of
systems containing only light quarks. The first is asymptotic
freedom, the fact that the effective coupling constant of QCD becomes
weak in processes with large momentum transfer, corresponding to
interactions at short-distance scales~\cite{Gros,Poli}. At large
distances, on the other hand, the coupling becomes strong, leading to
non-perturbative phenomena such as the confinement of quarks and
gluons on a length scale $R_{\rm had}\sim 1/\Lambda_{\rm QCD}\sim
1$~fm, which determines the size of hadrons~\cite{Maria}. Roughly
speaking, $\Lambda_{\rm QCD}\sim 0.2$ GeV is the energy scale that
separates the regions of large and small coupling constant. When the
mass of a quark $Q$ is much larger than this scale, it is called a
heavy quark. The quarks of the Standard Model fall naturally into two
classes: up, down and strange are light quarks, whereas charm, bottom
and top are heavy quarks.\footnote{Ironically, the top quark is of no
relevance to our discussion here, since it is too heavy to form
hadronic bound states before it decays.} 
For heavy quarks, the effective coupling constant $\alpha_s(m_Q)$ is
small, implying that on length scales comparable to the Compton
wavelength $\lambda_Q\sim 1/m_Q$ the strong interactions are
perturbative and similar to the electromagnetic interactions. In
fact, the quarkonium systems $(\bar QQ)$, whose size is of order
$\lambda_Q/\alpha_s(m_Q)\ll R_{\rm had}$, are very much
hydrogen-like.

Systems composed of a heavy quark and light constituents are
more complicated, however. The size of such systems is determined by
$R_{\rm had}$, and the typical momenta exchanged between the heavy
and light constituents are of order $\Lambda_{\rm QCD}$. The heavy
quark is surrounded by a most complicated, strongly interacting cloud
of light quarks, antiquarks, and gluons. In this case it is the fact
that $\lambda_Q\ll R_{\rm had}$, i.e.\ that the Compton wavelength of
the heavy quark is much smaller than the size of the hadron, which
leads to simplifications. To resolve the quantum numbers of the heavy
quark would require a hard probe; the soft gluons exchanged between
the heavy quark and the light constituents can only resolve distances
much larger than $\lambda_Q$. Therefore, the light degrees of freedom
are blind to the flavour (mass) and spin orientation of the heavy
quark. They experience only its colour field, which extends over
large distances because of confinement. In the rest frame of the
heavy quark, it is in fact only the electric colour field that is
important; relativistic effects such as colour magnetism vanish as
$m_Q\to\infty$. Since the heavy-quark spin participates in
interactions only through such relativistic effects, it decouples.
That the heavy-quark mass becomes irrelevant can be seen as follows:
As $m_Q\to\infty$, the heavy quark and the hadron that contains it
have the same velocity. In the rest frame of the hadron, the heavy
quark is at rest, too. The wave function of the light constituents
follows from a solution of the field equations of QCD subject to the
boundary condition of a static triplet source of colour at the
location of the heavy quark. This boundary condition is independent
of $m_Q$, and so is the solution for the configuration of the light
constituents.

It follows that, in the limit $m_Q\to\infty$, hadronic systems which
differ only in the flavour or spin quantum numbers of the heavy quark
have the same configuration of their light degrees of
freedom~\cite{Shu1}$^-$\cite{Isgu}. Although this observation still
does not allow us to calculate what this configuration is, it
provides relations between the properties of such particles as the
heavy mesons $B$, $D$, $B^*$ and $D^*$, or the heavy baryons
$\Lambda_b$ and $\Lambda_c$ (to the extent that corrections to the
infinite quark-mass limit are small in these systems). These
relations result from some approximate symmetries of the effective
strong interactions of heavy quarks at low energies. The
configuration of light degrees of freedom in a hadron containing a
single heavy quark with velocity $v$ does not change if this quark is
replaced by another heavy quark with different flavour or spin, but
with the same velocity. Both heavy quarks lead to the same static
colour field. For $N_h$ heavy-quark flavours, there is thus an SU$(2
N_h)$ spin--flavour symmetry group, under which the effective strong
interactions are invariant. These symmetries are in close
correspondence to familiar properties of atoms: The flavour symmetry
is analogous to the fact that different isotopes have the same
chemistry, since to a good approximation the wave function of the
electrons is independent of the mass of the nucleus. The electrons
only see the total nuclear charge. The spin symmetry is analogous to
the fact that the hyperfine levels in atoms are nearly degenerate.
The nuclear spin decouples in the limit $m_e/m_N\to 0$.

Heavy-quark symmetry is an approximate symmetry, and corrections
arise since the quark masses are not infinite. In many respects, it
is complementary to chiral symmetry, which arises in the opposite
limit of small quark masses. However, whereas chiral symmetry is a
symmetry of the QCD Lagrangian in the limit of vanishing quark
masses, heavy-quark symmetry is not a symmetry of the Lagrangian (not
even an approximate one), but rather a symmetry of an effective
theory, which is a good approximation of QCD in a certain kinematic
region. It is realized only in systems in which a heavy quark
interacts predominantly by the exchange of soft gluons. In such
systems the heavy quark is almost on shell; its momentum fluctuates
around the mass shell by an amount of order $\Lambda_{\rm QCD}$. The
corresponding fluctuations in the velocity of the heavy quark vanish
as $\Lambda_{\rm QCD}/m_Q\to 0$. The velocity becomes a conserved
quantity and is no longer a dynamical degree of freedom~\cite{Geor}.
Nevertheless, results derived on the basis of heavy-quark symmetry
are model-independent consequences of QCD in a well-defined limit.
The symmetry-breaking corrections can, at least in principle, be
studied in a systematic way. A convenient framework for analyzing
these corrections is provided by the heavy-quark effective theory.
Before presenting a detailed discussion of the formalism, we shall
first point out some of the important implications of heavy-quark
symmetry for the spectroscopy and weak decays of heavy hadrons.

\subsection{Spectroscopic Implications}

The spin--flavour symmetry leads to many interesting relations
between the properties of hadrons containing a heavy quark. The most
direct consequences concern the spectroscopy of such
states~\cite{IsWi}. In the limit $m_Q\to\infty$, the spin of the
heavy quark and the total angular momentum $j$ of the light degrees
of freedom inside a hadron are separately conserved by the strong
interactions. Because of heavy-quark symmetry, the dynamics is
independent of the spin and mass of the heavy quark. Hadronic states
can thus be classified by the quantum numbers (flavour, spin, parity,
etc.) of the light degrees of freedom~\cite{AFal}. The spin symmetry
predicts that, for fixed $j\neq 0$, there is a doublet of degenerate
states with total spin $J=j\pm\frac{1}{2}$. The flavour symmetry
relates the properties of states with different heavy-quark flavour.

In general, the mass of a hadron $H_Q$ containing a heavy quark $Q$
obeys an expansion of the form
\begin{equation}\label{massexp}
   m_H = m_Q + \bar\Lambda + {\Delta m^2\over 2 m_Q}
   + O(1/m_Q^2) \,.
\end{equation}
The parameter $\bar\Lambda$ represents contributions arising from all
terms in the Lagrangian that are independent of the heavy-quark
mass~\cite{FNL}, whereas the quantity $\Delta m^2$ originates from
the terms of order $1/m_Q$ in the effective Lagrangian of the HQET.
For the moment, the detailed structure of these terms is of no
relevance; it will be discussed at length in the next section. For
the ground-state pseudoscalar and vector mesons, one can parametrize
the contributions from the $1/m_Q$ corrections in terms of two
quantities, $\lambda_1$ and $\lambda_2$, in such a way
that~\cite{FaNe}
\begin{equation}\label{FNrela}
   \Delta m^2 = -\lambda_1 + 2 \Big[ J(J+1) - \textstyle{3\over 2}
   \Big]\,\lambda_2 \,.
\end{equation}
Here $J$ is the total spin of the meson. The first term,
$-\lambda_1/2 m_Q$, arises from the kinetic energy of the heavy quark
inside the meson; the second term describes the interaction of the
heavy-quark spin with the gluon field. The hadronic parameters
$\bar\Lambda$, $\lambda_1$ and $\lambda_2$ are independent of $m_Q$.
They characterize the properties of the light constituents.

Consider, as a first example, the SU(3) mass splittings for heavy
mesons. The heavy-quark expansion predicts that
\begin{eqnarray}
   m_{B_S} - m_{B_d} &=& \bar\Lambda_s - \bar\Lambda_d
    + O(1/m_b) \,, \nonumber\\
   m_{D_S} - m_{D_d} &=& \bar\Lambda_s - \bar\Lambda_d
    + O(1/m_c) \,,
\end{eqnarray}
where we have indicated that the value of the parameter $\bar\Lambda$
depends on the flavour of the light quark. Thus, to the extent that
the charm and bottom quarks can both be considered sufficiently
heavy, the mass splittings should be similar in the two systems. This
prediction is confirmed experimentally, since~\cite{Joe}
\begin{eqnarray}
   m_{B_S} - m_{B_d} &=& (90\pm 3)~\mbox{MeV} \,, \nonumber\\
   m_{D_S} - m_{D_d} &=& (99\pm 1)~\mbox{MeV} \,.
\end{eqnarray}

As a second example, consider the spin splittings between the
ground-state pseudoscalar ($J=0$) and vector ($J=1$) mesons, which
are the members of the spin-doublet with $j=\frac{1}{2}$. The theory
predicts that
\begin{eqnarray}
   m_{B^*}^2 - m_B^2 &=& 4\lambda_2 + O(1/m_b) \,, \nonumber\\
   m_{D^*}^2 - m_D^2 &=& 4\lambda_2 + O(1/m_c) \,.
\end{eqnarray}
The data are compatible with this:
\begin{eqnarray}\label{VPexp}
   m_{B^*}^2 - m_B^2 &\simeq& 0.49~{\rm GeV}^2 \,, \nonumber\\
   m_{D^*}^2 - m_D^2 &\simeq& 0.55~{\rm GeV}^2 \,.
\end{eqnarray}
Assuming that the $B$ system is close to the heavy-quark limit, we
obtain the value
\begin{equation}
   \lambda_2\simeq 0.12~\mbox{GeV}^2
\end{equation}
for one of the hadronic parameters in (\ref{FNrela}). This quantity
plays an important role in the phenomenology of inclusive decays of
heavy hadrons~\cite{review}.

A third example is provided by the mass splittings between the
ground-state mesons and baryons containing a heavy quark. The HQET
predicts that
\begin{eqnarray}\label{barmes}
   m_{\Lambda_b} - m_B &=& \bar\Lambda_{\rm baryon}
    - \bar\Lambda_{\rm meson} + O(1/m_b) \,, \nonumber\\
   m_{\Lambda_c} - m_D &=& \bar\Lambda_{\rm baryon}
    - \bar\Lambda_{\rm meson} + O(1/m_c) \,.
\end{eqnarray}
This is again consistent with the experimental results
\begin{eqnarray}
   m_{\Lambda_b} - m_B &=& (346\pm 6)~\mbox{MeV} \,, \nonumber\\
   m_{\Lambda_c} - m_D &=& (416\pm 1)~\mbox{MeV} \,,
\end{eqnarray}
although in this case the data indicate sizeable symmetry-breaking
corrections. For the mass of the $\Lambda_b$ baryon, we have used the
value
\begin{equation}\label{Lbmass}
   m_{\Lambda_b} = (5625\pm 6)~\mbox{MeV} \,,
\end{equation}
which is obtained by averaging the result~\cite{Joe} $m_{\Lambda_b}=
(5639\pm 15)$~MeV with the value $m_{\Lambda_b}=(5623\pm 5\pm 4)$~MeV
reported by the CDF Collaboration~\cite{CDFmass}. The dominant
correction to the relations (\ref{barmes}) comes from the
contribution of the chromo-magnetic interaction to the masses of the
heavy mesons,\footnote{Because of the spin symmetry, there is no such
contribution to the masses of the $\Lambda_Q$ baryons.} which adds a
term $3\lambda_2/2 m_Q$ on the right-hand side. Including this term,
we obtain the refined prediction that the values of the following two
quantities should be close to each other:
\begin{eqnarray}
   m_{\Lambda_b} - m_B - {3\lambda_2\over 2 m_B}
   &=& (312\pm 6)~\mbox{MeV} \,, \nonumber\\
   m_{\Lambda_c} - m_D - {3\lambda_2\over 2 m_D}
   &=& (320\pm 1)~\mbox{MeV}
\end{eqnarray}
This is clearly satisfied by the data.

The mass formula (\ref{massexp}) can also be used to derive
information on the heavy-quark (pole) masses from the observed hadron
masses. Introducing the ``spin-averaged'' meson masses
$\overline{m}_B=\frac{1}{4}\,(m_B+3 m_{B^*})\simeq 5.31$~GeV and
$\overline{m}_D=\frac{1}{4}\,(m_D+3 m_{D^*})\simeq 1.97$~GeV, we find
that
\begin{equation}\label{mbmc}
   m_b-m_c = (\overline{m}_B-\overline{m}_D)\,\bigg\{
   1 - {\lambda_1\over 2\overline{m}_B\overline{m}_D}
   + O(1/m_Q^3) \bigg\} \,,
\end{equation}
where $O(1/m_Q^3)$ is used as a generic notation representing terms
suppressed by three powers of the $b$- or $c$-quark masses. Using
theoretical estimates for the parameter $\lambda_1$, which lie in the
range~\cite{lam1}$^-$\cite{virial}
\begin{equation}\label{lam1}
   \lambda_1 = -(0.3\pm 0.2)~\mbox{GeV}^2 \,,
\end{equation}
this relation leads to
\begin{equation}\label{mbmcval}
   m_b - m_c = (3.39\pm 0.03\pm 0.03)~\mbox{GeV} \,,
\end{equation}
where the first error reflects the uncertainty in the value of
$\lambda_1$, and the second one takes into account unknown
higher-order corrections.

\subsection{Exclusive Semileptonic Decays}
\label{sec:3}

Semileptonic decays of $B$ mesons have received a lot of attention in
recent years. The decay channel $\bar B\to D^*\ell\,\bar\nu$ has the
largest branching fraction of all $B$-meson decay modes. From a
theoretical point of view, semileptonic decays are simple enough to
allow for a reliable, quantitative description. The analysis of these
decays provides much information about the strong forces that bind
the quarks and gluons into hadrons. Heavy-quark symmetry implies
relations between the weak decay form factors of heavy mesons, which
are of particular interest. These relations have been derived by
Isgur and Wise~\cite{Isgu}, generalizing ideas developed by Nussinov
and Wetzel~\cite{Nuss}, and by Voloshin and Shifman~\cite{Vol1,Vol2}.

\begin{figure}[htb]
   \epsfxsize=7cm
   \centerline{\epsffile{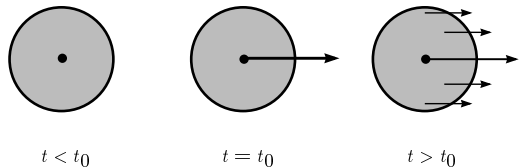}}
\caption{\label{fig:3.3}
Elastic transition induced by an external heavy-quark current.}
\end{figure}

Consider the elastic scattering of a $B$ meson, $\bar B(v)\to\bar
B(v')$, induced by a vector current coupled to the $b$ quark. Before
the action of the current, the light degrees of freedom inside the
$B$ meson orbit around the heavy quark, which acts as a static source
of colour. On average, the $b$ quark and the $B$ meson have the same
velocity $v$. The action of the current is to replace instantaneously
(at $t=t_0$) the colour source by one moving at a velocity $v'$, as
indicated in Fig.~\ref{fig:3.3}. If $v=v'$, nothing happens; the
light degrees of freedom do not realize that there was a current
acting on the heavy quark. If the velocities are different, however,
the light constituents suddenly find themselves interacting with a
moving colour source. Soft gluons have to be exchanged to rearrange
them so as to form a $B$ meson moving at velocity $v'$. This
rearrangement leads to a form-factor suppression, which reflects the
fact that as the velocities become more and more different, the
probability for an elastic transition decreases. The important
observation is that, in the limit $m_b\to\infty$, the form factor can
only depend on the Lorentz boost $\gamma = v\cdot v'$ that connects
the rest frames of the initial- and final-state mesons. Thus, in this
limit a dimensionless probability function $\xi(v\cdot v')$ describes
the transition. It is called the Isgur--Wise function~\cite{Isgu}. In
the HQET, which provides the appropriate framework for taking the
limit $m_b\to\infty$, the hadronic matrix element describing the
scattering process can thus be written as
\begin{equation}\label{elast}
   {1\over m_B}\,\langle\bar B(v')|\,\bar b_{v'}\gamma^\mu b_v\,
   |\bar B(v)\rangle = \xi(v\cdot v')\,(v+v')^\mu \,.
\end{equation}
Here, $b_v$ and $b_{v'}$ are the velocity-dependent heavy-quark
fields of the HQET, whose precise definition will be discussed in
Sec.~\ref{sec:HQET}. It is important that the function $\xi(v\cdot
v')$ does not depend on $m_b$. The factor $1/m_B$ on the left-hand
side compensates for a trivial dependence on the heavy-meson mass
caused by the relativistic normalization of meson states, which is
conventionally taken to be
\begin{equation}\label{nonrelnorm}
   \langle\bar B(p')|\bar B(p)\rangle = 2 m_B v^0\,(2\pi)^3\,
   \delta^3(\vec p-\vec p\,') \,.
\end{equation}
Note that there is no term proportional to $(v-v')^\mu$ in
(\ref{elast}). This can be seen by contracting the matrix element
with $(v-v')_\mu$, which must give zero since $\rlap/v b_v = b_v$ and
$\bar b_{v'}\rlap/v' = \bar b_{v'}$.

It is more conventional to write the above matrix element in terms of
an elastic form factor $F_{\rm el}(q^2)$ depending on the momentum
transfer $q^2=(p-p')^2$:
\begin{equation}
   \langle\bar B(v')|\,\bar b\,\gamma^\mu b\,|\bar B(v)\rangle
   = F_{\rm el}(q^2)\,(p+p')^\mu \,,
\end{equation}
where $p^(\phantom{}'\phantom{}^)=m_B v^(\phantom{}'\phantom{}^)$.
Comparing this with (\ref{elast}), we find that
\begin{equation}
   F_{\rm el}(q^2) = \xi(v\cdot v') \,, \qquad
   q^2 = -2 m_B^2 (v\cdot v'-1) \,.
\end{equation}
Because of current conservation, the elastic form factor is
normalized to unity at $q^2=0$. This condition implies the
normalization of the Isgur--Wise function at the kinematic point
$v\cdot v'=1$, i.e.\ for $v=v'$:
\begin{equation}\label{Jcons2}
   \xi(1) = 1 \,.
\end{equation}
It is in accordance with the intuitive argument that the probability
for an elastic transition is unity if there is no velocity change.
Since for $v=v'$ the daughter meson is at rest in the rest frame of
the parent meson, the point $v\cdot v'=1$ is referred to as the
zero-recoil limit.

We can now use the flavour symmetry to replace the $b$ quark in the
final-state meson by a $c$ quark, thereby turning the $B$ meson into
a $D$ meson. Then the scattering process turns into a weak decay
process. In the infinite mass limit, the replacement $b_{v'}\to
c_{v'}$ is a symmetry transformation, under which the effective
Lagrangian is invariant. Hence, the matrix element
\begin{equation}
   {1\over\sqrt{m_B m_D}}\,\langle D(v')|\,\bar c_{v'}\gamma^\mu
   b_v\,|\bar B(v)\rangle = \xi(v\cdot v')\,(v+v')^\mu
\end{equation}
is still determined by the same function $\xi(v\cdot v')$. This is
interesting, since in general the matrix element of a
flavour-changing current between two pseudoscalar mesons is described
by two form factors:
\begin{equation}
   \langle D(v')|\,\bar c\,\gamma^\mu b\,|\bar B(v)\rangle
   = f_+(q^2)\,(p+p')^\mu - f_-(q^2)\,(p-p')^\mu \,.
\end{equation}
Comparing the above two equations, we find that
\begin{eqnarray}\label{inelast}
   f_\pm(q^2) &=& {m_B\pm m_D\over 2\sqrt{m_B m_D}}\,\xi(v\cdot v')
    \,, \nonumber\\
   q^2 &=& m_B^2 + m_D^2 - 2 m_B m_D\,v\cdot v' \,.
\end{eqnarray}
Thus, the heavy-quark flavour symmetry relates two a priori
independent form factors to one and the same function. Moreover, the
normalization of the Isgur--Wise function at $v\cdot v'=1$ now
implies a non-trivial normalization of the form factors $f_\pm(q^2)$
at the point of maximum momentum transfer, $q_{\rm max}^2=
(m_B-m_D)^2$:
\begin{equation}
   f_\pm(q_{\rm max}^2) = {m_B\pm m_D\over 2\sqrt{m_B m_D}} \,.
\end{equation}

The heavy-quark spin symmetry leads to additional relations among
weak decay form factors. It can be used to relate matrix elements
involving vector mesons to those involving pseudoscalar mesons. A
vector meson with longitudinal polarization is related to a
pseudoscalar meson by a rotation of the heavy-quark spin. Hence, the
spin-symmetry transformation $c_{v'}^\Uparrow\to c_{v'}^\Downarrow$
relates $\bar B\to D$ with $\bar B\to D^*$ transitions. The result of
this transformation is~\cite{Isgu}:
\begin{eqnarray}
   {1\over\sqrt{m_B m_{D^*}}}\,
   \langle D^*(v',\varepsilon)|\,\bar c_{v'}\gamma^\mu b_v\,
   |\bar B(v)\rangle &=& i\epsilon^{\mu\nu\alpha\beta}\,
    \varepsilon_\nu^*\,v'_\alpha v_\beta\,\,\xi(v\cdot v') \,,
    \nonumber\\
   {1\over\sqrt{m_B m_{D^*}}}\,
   \langle D^*(v',\varepsilon)|\,\bar c_{v'}\gamma^\mu\gamma_5\,
   b_v\,|\bar B(v)\rangle &=& \Big[ \varepsilon^{*\mu}\,(v\cdot v'+1)
    - v'^\mu\,\varepsilon^*\!\cdot v \Big]\,\xi(v\cdot v') \,,
    \nonumber\\
\end{eqnarray}
where $\varepsilon$ denotes the polarization vector of the $D^*$ meson.
Once again, the matrix elements are completely described in terms of
the Isgur--Wise function. Now this is even more remarkable, since in
general four form factors, $V(q^2)$ for the vector current, and
$A_i(q^2)$, $i=0,1,2$, for the axial vector current, are required to
parametrize these matrix elements. In the heavy-quark limit, they
obey the relations~\cite{Neu1}
\begin{eqnarray}\label{PVff}
   {m_B\pm m_{D^*}\over 2\sqrt{m_B m_{D^*}}}\,\xi(v\cdot v')
   &=& V(q^2) = A_0(q^2) = A_1(q^2) \nonumber\\
   &=& \bigg[ 1 - {q^2\over(m_B+m_D)^2} \bigg]^{-1}\,A_1(q^2) \,,
    \nonumber\\
   \phantom{ \Bigg[ }
   q^2 &=& m_B^2 + m_{D^*}^2 - 2 m_B m_{D^*}\,v\cdot v' \,.
\end{eqnarray}

Equations (\ref{inelast}) and (\ref{PVff}) summarize the relations
imposed by heavy-quark symmetry on the weak decay form factors
describing the semileptonic decay processes $\bar B\to
D\,\ell\,\bar\nu$ and $\bar B\to D^*\ell\,\bar\nu$. These relations
are model-independent consequences of QCD in the limit where $m_b,
m_c\gg\Lambda_{\rm QCD}$. They play a crucial role in the
determination of the CKM matrix element $|V_{cb}|$. In terms of the
recoil variable $w=v\cdot v'$, the differential semileptonic decay
rates in the heavy-quark limit become~\cite{Vcb}:
\begin{eqnarray}\label{rates}
   {{\rm d}\Gamma(\bar B\to D\,\ell\,\bar\nu)\over{\rm d}w}
   &=& {G_F^2\over 48\pi^3}\,|V_{cb}|^2\,(m_B+m_D)^2\,m_D^3\,
    (w^2-1)^{3/2}\,\xi^2(w) \,, \nonumber\\
   \phantom{ \Bigg[ }
   {{\rm d}\Gamma(\bar B\to D^*\ell\,\bar\nu)\over{\rm d}w}
   &=& {G_F^2\over 48\pi^3}\,|V_{cb}|^2\,(m_B-m_{D^*})^2\,
    m_{D^*}^3\,\sqrt{w^2-1}\,(w+1)^2 \nonumber\\
   &&\times \Bigg[ 1 + {4w\over w+1}\,
    {m_B^2 - 2 w\,m_B m_{D^*} + m_{D^*}^2\over(m_B-m_{D^*})^2}
    \Bigg]\,\xi^2(w) \,.
\end{eqnarray}
These expressions receive symmetry-breaking corrections, since the
masses of the heavy quarks are not infinitely heavy. Perturbative
corrections of order $\alpha_s^n(m_Q)$ can be calculated order by
order in perturbation theory. A more difficult task is to control
the non-perturbative power corrections of order $(\Lambda_{\rm
QCD}/m_Q)^n$. The HQET provides a systematic framework for analysing
these corrections. For the case of weak-decay form factors, the
analysis of the $1/m_Q$ corrections was performed by
Luke~\cite{Luke}. Later, Falk and the present author have also
analysed the structure of $1/m_Q^2$ corrections for both meson and
baryon weak decay form factors~\cite{FaNe}. We shall not discuss
these rather technical issues in detail, but only mention the most
important result of Luke's analysis. It concerns the zero-recoil
limit, where an analogue of the Ademollo--Gatto theorem~\cite{AGTh}
can be proved. This is Luke's theorem~\cite{Luke}, which states that
the matrix elements describing the leading $1/m_Q$ corrections to
weak decay amplitudes vanish at zero recoil. This theorem is valid to
all orders in perturbation theory~\cite{FaNe,Neu7,ChGr}. Most
importantly, it protects the $\bar B\to D^*\ell\,\bar\nu$ decay rate
from receiving first-order $1/m_Q$ corrections at zero
recoil~\cite{Vcb}. (A similar statement is not true for the decay
$\bar B\to D\,\ell\,\bar\nu$, however. The reason is simple but
somewhat subtle. Luke's theorem protects only those form factors not
multiplied by kinematic factors that vanish for $v=v'$. By angular
momentum conservation, the two pseudoscalar mesons in the decay $\bar
B\to D\,\ell\,\bar\nu$ must be in a relative $p$ wave, and hence the
amplitude is proportional to the velocity $|\vec v_D|$ of the $D$
meson in the $B$-meson rest frame. This leads to a factor $(w^2-1)$
in the decay rate. In such a situation, form factors that are
kinematically suppressed can contribute~\cite{Neu1}.)

\subsection{Model-Independent Determination of $|V_{cb}|$}
 
We will now discuss the most
important application of the HQET in the context of semileptonic
decays of $B$ mesons. A model-independent determination of the CKM
matrix element $|V_{cb}|$ based on heavy-quark symmetry can be
obtained by measuring the recoil spectrum of $D^*$ mesons produced in
$\bar B\to D^*\ell\,\bar\nu$ decays~\cite{Vcb}. In the heavy-quark
limit, the differential decay rate for this process has been given in
(\ref{rates}). In order to allow for corrections to that limit, we
write
\begin{eqnarray}
   {{\rm d}\Gamma(\bar B\to D^*\ell\,\bar\nu)\over{\rm d}w}
   &=& {G_F^2\over 48\pi^3}\,(m_B-m_{D^*})^2\,m_{D^*}^3
    \sqrt{w^2-1}\,(w+1)^2 \nonumber\\
   &&\mbox{}\times \Bigg[ 1 + {4w\over w+1}\,
    {m_B^2-2w\,m_B m_{D^*} + m_{D^*}^2\over(m_B - m_{D^*})^2}
    \Bigg]\,|V_{cb}|^2\,{\cal{F}}^2(w) \,, \nonumber\\
\end{eqnarray}
where the hadronic form factor ${\cal F}(w)$ coincides with the
Isgur--Wise function up to symmetry-breaking corrections of order
$\alpha_s(m_Q)$ and $\Lambda_{\rm QCD}/m_Q$. The idea is to measure
the product $|V_{cb}|\,{\cal F}(w)$ as a function of $w$, and to
extract $|V_{cb}|$ from an extrapolation of the data to the
zero-recoil point $w=1$, where the $B$ and the $D^*$ mesons have a
common rest frame. At this kinematic point, heavy-quark symmetry
helps to calculate the normalization ${\cal F}(1)$ with small and
controlled theoretical errors. Since the range of $w$ values
accessible in this decay is rather small ($1<w<1.5$), the
extrapolation can be done using an expansion around $w=1$:
\begin{equation}\label{Fexp}
   {\cal F}(w) = {\cal F}(1)\,\Big[ 1 - \widehat\varrho^2\,(w-1)
   + \dots \Big] \,.
\end{equation}
The slope $\widehat\varrho^2$ is treated as a fit parameter.

\begin{figure}[htb]
   \epsfxsize=8cm
   \vspace{0.3cm}
   \centerline{\epsffile{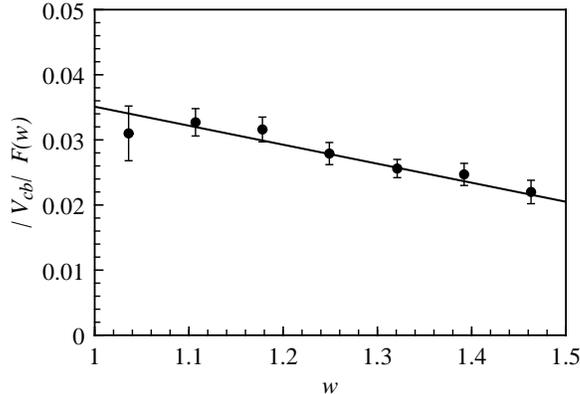}}
   \vspace{-0.3cm}
\caption{\label{fig:CLVcb}
CLEO data for the product $|V_{cb}|\,{\cal F}(w)$, as extracted from
the recoil spectrum in $\bar B\to D^*\ell\,\bar\nu$
decays~\protect\cite{CLEOVcb}. The line shows a linear fit to the
data.}
\end{figure}

Measurements of the recoil spectrum have been performed first by the
ARGUS~\cite{ARGVcb} and CLEO~\cite{CLEOVcb} Collaborations in
experiments operating at the $\Upsilon(4s)$ resonance, and more
recently by the ALEPH~\cite{ALEVcb} and DELPHI~\cite{DELVcb}
Collaborations at LEP. As an example, Fig.~\ref{fig:CLVcb} shows the
data reported by the CLEO Collaboration. The results obtained by the
various experimental groups from a linear fit to their data are
summarized in Table~\ref{tab:Vcb}. The weighted average of these
results is
\begin{eqnarray}\label{VcbFraw}
   |V_{cb}|\,{\cal F}(1) &=& (34.6\pm 1.7)\times 10^{-3} \,,
    \nonumber\\
   \widehat\varrho^2 &=& 0.82\pm 0.09 \,.
\end{eqnarray}
The effect of a positive curvature of the form factor has been
investigated by Stone~\cite{Stone}, who finds that the value of
$|V_{cb}|\,{\cal F}(1)$ may change by up to $+4\%$. We thus increase
the above value by $(2\pm 2)\%$ and quote the final result as
\begin{equation}\label{VcbF}
   |V_{cb}|\,{\cal F}(1) = (35.3\pm 1.8)\times 10^{-3} \,.
\end{equation}
In future analyses, the extrapolation to zero recoil should be
performed including higher-order terms in the expansion (\ref{Fexp}).
It can be shown in a model-independent way that the shape of the form
factor is highly constrained by analyticity and unitarity
requirements~\cite{Boyd2,Capr}. In particular, the curvature at $w=1$
is strongly correlated with the slope of the form factor. For the
value of $\widehat\varrho^2$ given in (\ref{VcbFraw}), one obtains a
small positive curvature~\cite{Capr}, in agreement with the
assumption made in Ref.~47.

\begin{table}[htb]
\caption{\label{tab:Vcb}
Values for $|V_{cb}|\,{\cal F}(1)$ (in units of $10^{-3}$) and
$\widehat\varrho^2$ extracted from measurements of the recoil
spectrum in $\bar B\to D^*\ell\,\bar\nu$ decays}
\vspace{0.4cm}
\begin{center}
\begin{tabular}{|l|c|c|}\hline
\rule[-0.15cm]{0cm}{0.65cm} & $|V_{cb}|\,{\cal F}(1)~(10^{-3})$ &
 $\widehat\varrho^2$ \\
\hline
ARGUS  & $38.8\pm 4.3\pm 2.5$ & $1.17\pm 0.22\pm 0.06$ \\
CLEO   & $35.1\pm 1.9\pm 2.0$ & $0.84\pm 0.12\pm 0.08$ \\
ALEPH  & $31.4\pm 2.3\pm 2.5$ & $0.39\pm 0.21\pm 0.12$ \\
DELPHI & $35.0\pm 1.9\pm 2.3$ & $0.81\pm 0.16\pm 0.10$ \\
\hline
\end{tabular}
\end{center}
\end{table}

Heavy-quark symmetry implies that the general structure of the
symmetry-breaking corrections to the form factor at zero recoil
is~\cite{Vcb}
\begin{equation}
   {\cal F}(1) = \eta_A\,\bigg( 1 + 0 \times
   {\Lambda_{\rm QCD}\over m_Q}
   + \mbox{const} \times {\Lambda_{\rm QCD}^2\over m_Q^2}
   + \dots \bigg)
   \equiv \eta_A\,(1+\delta_{1/m^2}) \,,
\end{equation}
where $\eta_A$ is a short-distance correction arising from the
(finite) renormalization of the flavour-changing axial current at
zero recoil, and $\delta_{1/m^2}$ parametrizes second-order (and
higher) power corrections. The absence of first-order power
corrections at zero recoil is a consequence of Luke's
theorem~\cite{Luke}. The one-loop expression for $\eta_A$ has been
known for a long time~\cite{Pasc,Vol2,QCD1}:
\begin{equation}\label{etaA1}
   \eta_A = 1 + {\alpha_s(M)\over\pi}\,\bigg(
   {m_b+m_c\over m_b-m_c}\,\ln{m_b\over m_c} - {8\over 3} \bigg)
   \simeq 0.96 \,.
\end{equation}
The scale $M$ in the running coupling constant can be fixed by
adopting the prescription of Brodsky, Lepage and Mackenzie
(BLM)~\cite{BLM}, according to which it is identified with the
average virtuality of the gluon in the one-loop diagrams that
contribute to $\eta_A$. If $\alpha_s(M)$ is defined in the modified
minimal subtraction ($\overline{\mbox{\sc ms}}$) scheme, the result
is~\cite{etaVA} $M\simeq 0.51\sqrt{m_c m_b}$. Several estimates of
higher-order corrections to $\eta_A$ have been discussed. The
next-to-leading order resummation of logarithms of the type
$[\alpha_s\ln( m_b/m_c)]^n$ leads to~\cite{FaGr,QCD2} $\eta_A\simeq
0.985$. On the other hand, the resummation of ``renormalon-chain''
contributions of the form $\beta_0^{n-1}\alpha_s^n$, where $\beta_0$
is the first coefficient of the QCD $\beta$-function,
gives~\cite{flow} $\eta_A\simeq 0.945$. Using these partial
resummations to estimate the uncertainty results in $\eta_A =
0.965\pm 0.020$. Recently, Czarnecki has improved this estimate by
calculating $\eta_A$ at two-loop order~\cite{Czar}. His result,
\begin{equation}
   \eta_A = 0.960\pm 0.007 \,,
\end{equation}
is in excellent agreement with the BLM-improved one-loop estimate
(\ref{etaA1}). Here the error is taken to be the size of the two-loop
correction.

The analysis of the power corrections $\delta_{1/m^2}$ is more
difficult, since it cannot rely on perturbation theory. Three
approaches have been discussed: in the ``exclusive approach'', all
$1/m_Q^2$ operators in the HQET are classified and their matrix
elements estimated, leading to~\cite{FaNe,TMann}
$\delta_{1/m^2}=-(3\pm 2)\%$; the ``inclusive approach'' has been
used to derive the bound $\delta_{1/m^2}<-3\%$, and to estimate
that~\cite{Shif}$^,$\footnote{This bound has been criticised in
Ref.~59.}
$\delta_{1/m^2}=-(7\pm 3)\%$; the ``hybrid approach'' combines the
virtues of the former two to obtain a more restrictive lower bound on
$\delta_{1/m^2}$. This leads to~\cite{Vcbnew}
\begin{equation}
   \delta_{1/m^2} = - 0.055\pm 0.025 \,.
\end{equation}

Combining the above results, adding the theoretical errors linearly
to be conservative, gives
\begin{equation}\label{F1}
   {\cal F}(1) = 0.91\pm 0.03
\end{equation}
for the normalization of the hadronic form factor at zero recoil.
Thus, the corrections to the heavy-quark limit amount to a moderate
decrease of the form factor of about 10\%. This can be used to
extract from the experimental result (\ref{VcbF}) the
model-independent value
\begin{equation}\label{Vcbexc}
   |V_{cb}| = (38.8\pm 2.0_{\rm exp}\pm 1.2_{\rm th})
   \times 10^{-3} \,.
\end{equation}

\section{Heavy-Quark Effective Theory}
\label{sec:HQET}

\subsection{The Effective Lagrangian}
\label{sec:Leff}

The effects of a very heavy particle often become irrelevant at low
energies. It is then useful to construct a low-energy effective
theory, in which this heavy particle no longer appears. Eventually,
this effective theory will be easier to deal with than the full
theory. A familiar example is Fermi's theory of the weak
interactions. For the description of weak decays of hadrons, the weak
interactions can be approximated by point-like four-fermion
couplings, governed by a dimensionful coupling constant $G_F$. Only
at energies much larger than the masses of hadrons can the effects of
the intermediate vector bosons, $W$ and $Z$, be resolved.

The process of removing the degrees of freedom of a heavy particle
involves the following steps~\cite{SVZ1}$^-$\cite{Polc}: one first
identifies the heavy-particle fields and ``integrates them out'' in
the generating functional of the Green functions of the theory. This
is possible since at low energies the heavy particle does not appear
as an external state. However, although the action of the full theory
is usually a local one, what results after this first step is a
non-local effective action. The non-locality is related to the fact
that in the full theory the heavy particle with mass $M$ can appear
in virtual processes and propagate over a short but finite distance
$\Delta x\sim 1/M$. Thus, a second step is required to obtain a local
effective Lagrangian: the non-local effective action is rewritten as
an infinite series of local terms in an Operator Product Expansion
(OPE)~\cite{Wils,Zimm}. Roughly speaking, this corresponds to an
expansion in powers of $1/M$. It is in this step that the short- and
long-distance physics is disentangled. The long-distance physics
corresponds to interactions at low energies and is the same in the
full and the effective theory. But short-distance effects arising
from quantum corrections involving large virtual momenta (of order
$M$) are not reproduced in the effective theory, once the heavy
particle has been integrated out. In a third step, they have to be
added in a perturbative way using renormalization-group techniques.
These short-distance effects lead to a renormalization of the
coefficients of the local operators in the effective Lagrangian. An
example is the effective Lagrangian for non-leptonic weak decays, in
which radiative corrections from hard gluons with virtual momenta in
the range between $m_W$ and some renormalization scale $\mu\sim
1$~GeV give rise to Wilson coefficients, which renormalize the local
four-fermion interactions~\cite{AltM}$^-$\cite{Gilm}.

\begin{figure}[htb]
   \epsfysize=8cm
   \centerline{\epsffile{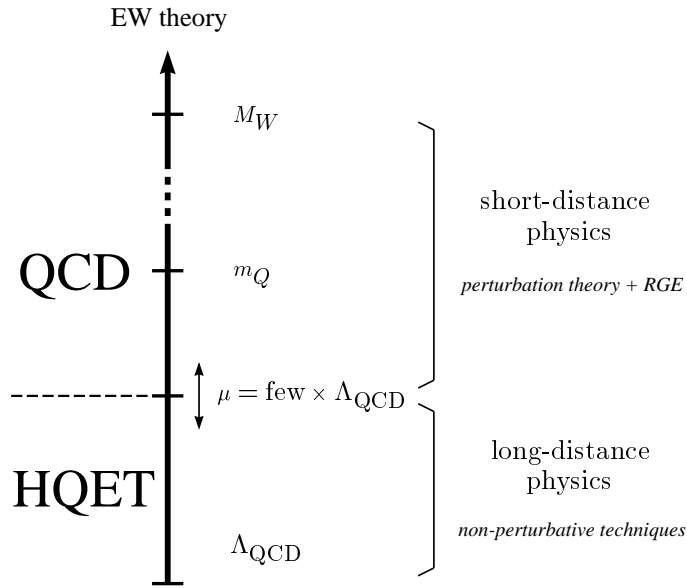}}
\caption{\label{fig:magic}
Philosophy of the heavy-quark effective theory.}
\end{figure}

The heavy-quark effective theory (HQET) is constructed to provide a
simplified description of processes where a heavy quark interacts
with light degrees of freedom predominantly by the exchange of soft
gluons~\cite{EiFe}$^-$\cite{Mann}. Clearly, $m_Q$ is the high-energy
scale in this case, and $\Lambda_{\rm QCD}$ is the scale of the
hadronic physics we are interested in. The situation is illustrated
in Fig.~\ref{fig:magic}. At short distances, i.e.\ for energy scales
larger than the heavy-quark mass, the physics is perturbative and
described by ordinary QCD. For mass scales much below the heavy-quark
mass, the physics is complicated and non-perturbative because of
confinement. Our goal is to obtain a simplified description in this
region using an effective field theory. To separate short- and
long-distance effects, we introduce a separation scale $\mu$ such
that $\Lambda_{\rm QCD}\ll\mu\ll m_Q$. The HQET will be constructed
in such a way that it is identical to QCD in the long-distance
region, i.e.\ for scales below $\mu$. In the short-distance region,
the effective theory is incomplete, however, since some high-momentum
modes have been integrated out from the full theory. The fact that
the physics must be independent of the arbitrary scale $\mu$ allows
us to derive renormalization-group equations, which we shall employ
to deal with the short-distance effects in an efficient way.

Compared with most effective theories, in which the degrees of
freedom of a heavy particle are removed completely from the
low-energy theory, the HQET is special in that its purpose is to
describe the properties and decays of hadrons which do contain a
heavy quark. Hence, it is not possible to remove the heavy quark
completely from the effective theory. What is possible is to
integrate out the ``small components'' in the full heavy-quark
spinor, which describe the fluctuations around the mass shell.

The starting point in the construction of the low-energy effective
theory is the observation that a very heavy quark bound inside a
hadron moves more or less with the hadron's velocity $v$, and is
almost on shell. Its momentum can be written as
\begin{equation}\label{kresdef}
   p_Q^\mu = m_Q v^\mu + k^\mu \,,
\end{equation}
where the components of the so-called residual momentum $k$ are much
smaller than $m_Q$. Note that $v$ is a four-velocity, so that
$v^2=1$. Interactions of the heavy quark with light degrees of
freedom change the residual momentum by an amount of order $\Delta
k\sim\Lambda_{\rm QCD}$, but the corresponding changes in the
heavy-quark velocity vanish as $\Lambda_{\rm QCD}/m_Q\to 0$. In this
situation, it is appropriate to introduce large- and small-component
fields, $h_v$ and $H_v$, by
\begin{equation}\label{hvHvdef}
   h_v(x) = e^{i m_Q v\cdot x}\,P_+\,Q(x) \,, \qquad
   H_v(x) = e^{i m_Q v\cdot x}\,P_-\,Q(x) \,,
\end{equation}
where $P_+$ and $P_-$ are projection operators defined as
\begin{equation}
   P_\pm = {1\pm\rlap/v\over 2} \,.
\end{equation}
It follows that
\begin{equation}\label{redef}
   Q(x) = e^{-i m_Q v\cdot x}\,[ h_v(x) + H_v(x) ] \,.
\end{equation}
Because of the projection operators, the new fields satisfy
$\rlap/v\,h_v=h_v$ and $\rlap/v\,H_v=-H_v$. In the rest frame, i.e.\
for $v^\mu=(1,0,0,0)$, $h_v$ corresponds to the upper two components
of $Q$, while $H_v$ corresponds to the lower ones. Whereas $h_v$
annihilates a heavy quark with velocity $v$, $H_v$ creates a heavy
antiquark with velocity $v$.

\begin{figure}[htb]
   \epsfxsize=7cm
   \centerline{\epsffile{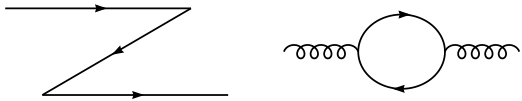}}
\caption{\label{fig:3.1}
Virtual fluctuations involving pair creation of heavy quarks. In the
first diagram, time flows to the right.}
\end{figure}

In terms of the new fields, the QCD Lagrangian for a heavy quark
takes the form
\begin{eqnarray}\label{Lhchi}
   {\cal L}_Q &=& \bar Q\,(i\,\rlap{\,/}D - m_Q)\,Q \nonumber\\
   &=& \bar h_v\,i v\!\cdot\!D\,h_v
    - \bar H_v\,(i v\!\cdot\!D + 2 m_Q)\,H_v \nonumber\\
   &&\mbox{}+ \bar h_v\,i\,\rlap{\,/}D_\perp H_v
    + \bar H_v\,i\,\rlap{\,/}D_\perp h_v \,,
\end{eqnarray}
where $D_\perp^\mu = D^\mu - v^\mu\,v\cdot D$ is orthogonal to the
heavy-quark velocity: $v\cdot D_\perp=0$. In the rest frame,
$D_\perp^\mu=(0,\vec D\,)$ contains the spatial components of the
covariant derivative. From (\ref{Lhchi}), it is apparent that $h_v$
describes massless degrees of freedom, whereas $H_v$ corresponds to
fluctuations with twice the heavy-quark mass. These are the heavy
degrees of freedom that will be eliminated in the construction of the
effective theory. The fields are mixed by the presence of the third
and fourth terms, which describe pair creation or annihilation of
heavy quarks and antiquarks. As shown in the first diagram in
Fig.~\ref{fig:3.1}, in a virtual process a heavy quark propagating
forward in time can turn into an antiquark propagating backward in
time, and then turn back into a quark. The energy of the intermediate
quantum state $h h\bar H$ is larger than the energy of the initial
heavy quark by at least $2 m_Q$. Because of this large energy gap,
the virtual quantum fluctuation can only propagate over a short
distance $\Delta x\sim 1/m_Q$. On hadronic scales set by $R_{\rm
had}=1/\Lambda_{\rm QCD}$, the process essentially looks like a local
interaction of the form
\begin{equation}
   \bar h_v\,i\,\rlap{\,/}D_\perp\,{1\over 2 m_Q}\,
   i\,\rlap{\,/}D_\perp h_v \,,
\end{equation}
where we have simply replaced the propagator for $H_v$ by $1/2 m_Q$.
A more correct treatment is to integrate out the small-component
field $H_v$, thereby deriving a non-local effective action for the
large-component field $h_v$, which can then be expanded in terms of
local operators. Before doing this, let us mention a second type of
virtual corrections involving pair creation, namely heavy-quark
loops. An example is shown in the second diagram in
Fig.~\ref{fig:3.1}. Heavy-quark loops cannot be described in terms of
the effective fields $h_v$ and $H_v$, since the quark velocities
inside a loop are not conserved and are in no way related to hadron
velocities. However, such short-distance processes are proportional
to the small coupling constant $\alpha_s(m_Q)$ and can be calculated
in perturbation theory. They lead to corrections that are added onto
the low-energy effective theory in the renormalization procedure to
be discussed later.

On a classical level, the heavy degrees of freedom represented by
$H_v$ can be eliminated using the equation of motion. Taking the
variation of the Lagrangian with respect to the field $\bar H_v$, we
obtain
\begin{equation}
   (i v\!\cdot\!D + 2 m_Q)\,H_v = i\,\rlap{\,/}D_\perp h_v \,.
\end{equation}
This equation can formally be solved to give
\begin{equation}\label{Hfield}
   H_v = {1\over 2 m_Q + i v\!\cdot\!D}\,
   i\,\rlap{\,/}D_\perp h_v \,,
\end{equation}
showing that the small-component field $H_v$ is indeed of order
$1/m_Q$. We can now insert this solution into (\ref{Lhchi}) to obtain
the ``non-local effective Lagrangian''
\begin{equation}\label{Lnonloc}
   {\cal L}_{\rm eff} = \bar h_v\,i v\!\cdot\!D\,h_v
   + \bar h_v\,i\,\rlap{\,/}D_\perp\,{1\over 2 m_Q+i v\!\cdot\!D}\,
   i\,\rlap{\,/}D_\perp h_v \,.
\end{equation}
Clearly, the second term corresponds to the first class of virtual
processes shown in Fig.~\ref{fig:3.1}.

It is possible to derive this Lagrangian in a more elegant way by
manipulating the generating functional for QCD Green's functions
containing heavy-quark fields~\cite{Mann}. To this end, one starts
from the field redefinition (\ref{redef}) and couples the
large-component fields $h_v$ to external sources $\rho_v$. Green's
functions with an arbitrary number of $h_v$ fields can be constructed
by taking derivatives with respect to $\rho_v$. No sources are needed
for the heavy degrees of freedom represented by $H_v$. The functional
integral over these fields is Gaussian and can be performed
explicitly, leading to the effective action
\begin{equation}\label{SeffMRR}
   S_{\rm eff} = \int\!{\rm d}^4 x\,{\cal L}_{\rm eff}
   - i \ln\Delta \,,
\end{equation}
with ${\cal L}_{\rm eff}$ as given in (\ref{Lnonloc}). The
appearance
of the logarithm of the determinant
\begin{equation}
   \Delta = \exp\bigg( {1\over 2}\,{\rm Tr}\,
   \ln\big[ 2 m_Q + i v\!\cdot\!D - i\eta \big] \bigg)
\end{equation}
is a quantum effect not present in the classical derivation presented
above. However, in this case the determinant can be regulated in a
gauge-invariant way, and by choosing the axial gauge $v\cdot A=0$ one
shows that $\ln\Delta$ is just an irrelevant
constant~\cite{Mann,Soto}.

\begin{figure}[htb]
   \epsfysize=3cm
   \centerline{\epsffile{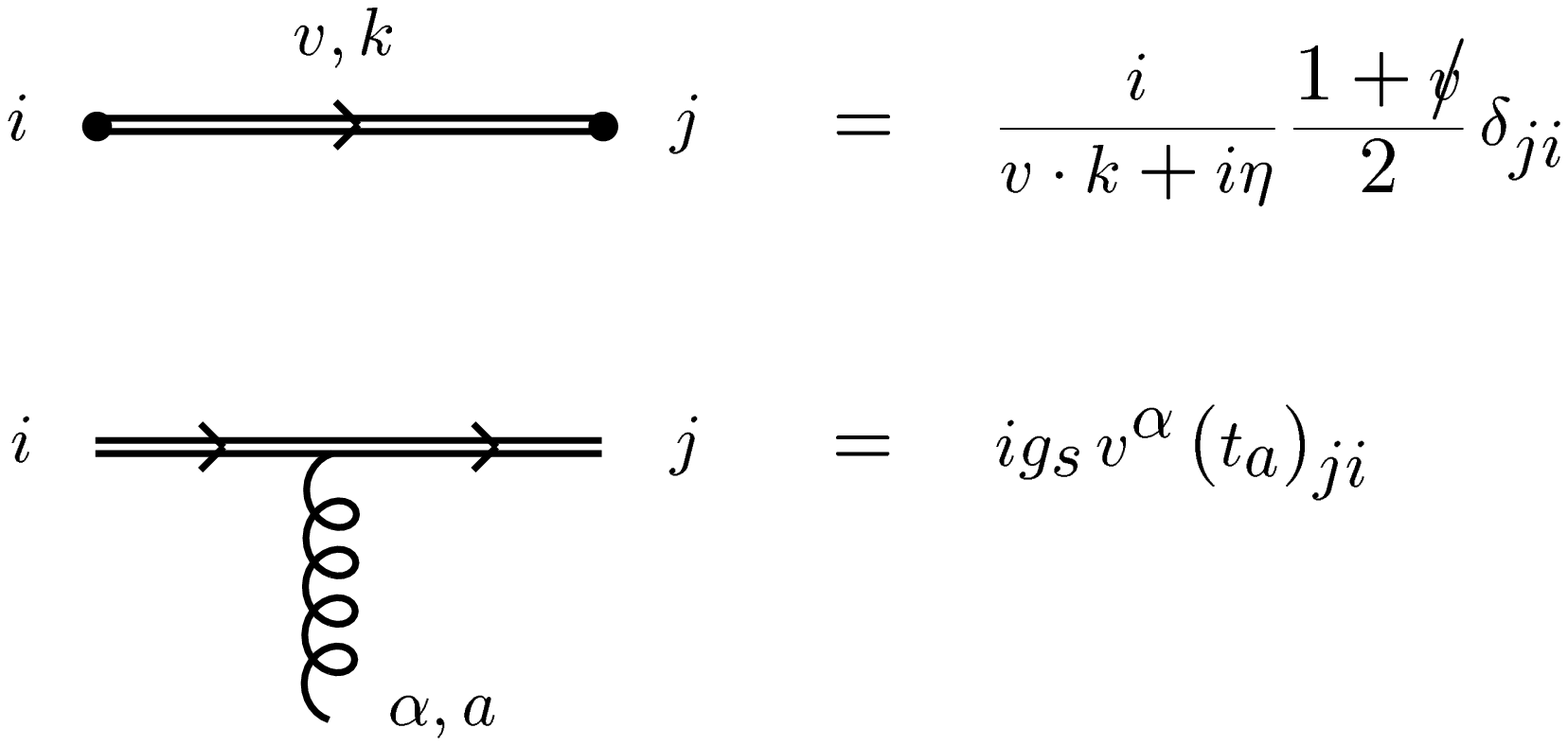}}
\caption{\label{fig:3.2}
Feynman rules of the HQET ($i,j$ and $a$ are colour indices). A heavy
quark is represented by a double line labelled by the velocity $v$
and the residual momentum $k$. The velocity $v$ is conserved by the
strong interactions.}
\end{figure}

Because of the phase factor in (\ref{redef}), the $x$ dependence of
the effective heavy-quark field $h_v$ is weak. In momentum space,
derivatives acting on $h_v$ correspond to powers of the residual
momentum $k$, which by construction is much smaller than $m_Q$.
Hence, the non-local effective Lagrangian (\ref{Lnonloc}) allows for
a derivative expansion in powers of $iD/m_Q$:
\begin{equation}
   {\cal L}_{\rm eff} = \bar h_v\,i v\!\cdot\!D\,h_v
   + {1\over 2 m_Q}\,\sum_{n=0}^\infty\,
   \bar h_v\,i\,\rlap{\,/}D_\perp\,\bigg( -{i v\cdot D\over 2 m_Q}
   \bigg)^n\,i\,\rlap{\,/}D_\perp h_v \,.
\end{equation}
Taking into account that $h_v$ contains a $P_+$ projection operator,
and using the identity
\begin{equation}\label{pplusid}
   P_+\,i\,\rlap{\,/}D_\perp\,i\,\rlap{\,/}D_\perp P_+
   = P_+\,\bigg[ (i D_\perp)^2 + {g_s\over 2}\,
   \sigma_{\mu\nu }\,G^{\mu\nu } \bigg]\,P_+ \,,
\end{equation}
where $[i D^\mu,i D^\nu]=i g_s G^{\mu\nu}$ is the gluon
field-strength tensor, one finds that~\cite{EiH2,FGL}
\begin{equation}\label{Lsubl}
   {\cal L}_{\rm eff} = \bar h_v\,i v\!\cdot\!D\,h_v
   + {1\over 2 m_Q}\,\bar h_v\,(i D_\perp)^2\,h_v
   + {g_s\over 4 m_Q}\,\bar h_v\,\sigma_{\mu\nu}\,
   G^{\mu\nu}\,h_v + O(1/m_Q^2) \,.
\end{equation}
In the limit $m_Q\to\infty$, only the first terms remains:
\begin{equation}\label{Leff}
   {\cal L}_\infty = \bar h_v\,i v\!\cdot\!D\,h_v \,.
\end{equation}
This is the effective Lagrangian of the HQET. It gives rise to the
Feynman rules depicted in Fig.~\ref{fig:3.2}.

Let us take a moment to study the symmetries of this
Lagrangian~\cite{Geor}. Since there appear no Dirac matrices,
interactions of the heavy quark with gluons leave its spin unchanged.
Associated with this is an SU(2) symmetry group, under which
${\cal L}_\infty$ is invariant. The action of this symmetry on the
heavy-quark fields becomes most transparent in the rest frame, where
the generators $S^i$ of SU(2) can be chosen as
\begin{equation}\label{Si}
   S^i = {1\over 2} \left( \begin{array}{cc}
                           \sigma^i ~&~ 0 \\
                           0 ~&~ \sigma^i \end{array} \right) \,,
   \qquad [S^i,S^j] = i \epsilon^{ijk} S^k \,.
\end{equation}
Here $\sigma^i$ are the Pauli matrices. An infinitesimal SU(2)
transformation $h_v\to (1 + i\vec\epsilon \cdot\vec S\,)\,h_v$ leaves
the Lagrangian invariant:
\begin{equation}\label{SU2tr}
   \delta{\cal L}_\infty = \bar h_v\,
   [i v\!\cdot\! D,i \vec\epsilon\cdot\vec S\,]\,h_v = 0 \,.
\end{equation}
Another symmetry of the HQET arises since the mass of the heavy quark
does not appear in the effective Lagrangian. For $N_h$ heavy quarks
moving at the same velocity, eq.~(\ref{Leff}) can be extended by
writing
\begin{equation}\label{Leff2}
   {\cal L}_\infty
   = \sum_{i=1}^{N_h}\,\bar h_v^i\,i v\!\cdot\! D\,h_v^i \,.
\end{equation}
This is invariant under rotations in flavour space. When combined
with the spin symmetry, the symmetry group is promoted to SU$(2N_h)$.
This is the heavy-quark spin--flavour symmetry~\cite{Isgu,Geor}. Its
physical content is that, in the limit $m_Q\to\infty$, the strong
interactions of a heavy quark become independent of its mass and
spin.

Consider now the operators appearing at order $1/m_Q$ in the
effective Lagrangian (\ref{Lsubl}). They are easiest to identify in
the rest frame. The first operator,
\begin{equation}\label{Okin}
   {\cal O}_{\rm kin} = {1\over 2 m_Q}\,\bar h_v\,(i D_\perp)^2\,
   h_v \to - {1\over 2 m_Q}\,\bar h_v\,(i \vec D\,)^2\,h_v \,,
\end{equation}
is the gauge-covariant extension of the kinetic energy arising from
the off-shell residual motion of the heavy quark. The second operator
is the non-abelian analogue of the Pauli interaction, which describes
the chromo-magnetic coupling of the heavy-quark spin to the gluon
field:
\begin{equation}\label{Omag}
   {\cal O}_{\rm mag} = {g_s\over 4 m_Q}\,\bar h_v\,
   \sigma_{\mu\nu}\,G^{\mu\nu}\,h_v \to
   - {g_s\over m_Q}\,\bar h_v\,\vec S\!\cdot\!\vec B_c\,h_v \,.
\end{equation}
Here $\vec S$ is the spin operator defined in (\ref{Si}), and $B_c^i
= -\frac{1}{2}\epsilon^{ijk} G^{jk}$ are the components of the
chromo-magnetic field. The chromo-magnetic interaction is a
relativistic effect, which scales like $1/m_Q$. This is the origin of
the heavy-quark spin symmetry.

\subsection{Wave-Function Renormalization of the Heavy-Quark Field in
the HQET}

Besides being an effective theory for the strong interactions of
heavy quarks with light degrees of freedom, the HQET is a consistent,
renormalizable (order by order in $1/m_Q$) quantum field theory in
its own right. In particular, it provides a framework for calculating
radiative corrections. We shall discuss as an illustration the
wave-function renormalization of the heavy-quark field $h_v$.

In quantum field theory, the parameters and fields of the Lagrangian
have no direct physical significance. They have to be renormalized
before they can be related to observable quantities. In an
intermediate step the theory has to be regularized. The most
convenient regularization scheme in QCD is dimensional
regularization~\cite{tHo2}$^-$\cite{Boll}, in which the dimension of
space-time is analytically continued to $D=4-2\epsilon$, with
$\epsilon$ being infinitesimal. Loop integrals that are
logarithmically divergent in four dimensions become finite for
$\epsilon>0$. From the fact that the action $S=\int{\rm d}^Dx\, {\cal
L}(x)$ is dimensionless, one can derive the mass dimensions of the
fields and parameters of the theory. For instance, one finds that the
``bare'' coupling constant $\alpha_s^{\rm bare}$ is no longer
dimensionless if $D\ne 4$: ${\rm dim}[\,\alpha_s^{\rm bare}\,] =
2\epsilon$. In a renormalizable theory, it is possible to rewrite the
Lagrangian in terms of renormalized quantities in such a way that
Green's functions of the renormalized fields remain finite as
$\epsilon\to 0$. For QCD, one introduces renormalized quantities by
$Q^{\rm bare} = Z_Q^{1/2}\,Q^{\rm ren}$, $A^{\rm bare} = Z_A^{1/2}
A^{\rm ren}$, $\alpha_s^{\rm bare}=\mu^{2\epsilon}Z_\alpha\,
\alpha_s^{\rm ren}$, etc., where $\mu$ is an arbitrary mass scale
introduced to render the renormalized coupling constant
dimensionless. Similarly, in the HQET one defines the renormalized
heavy-quark field by $h_v^{\rm bare}=Z_h^{1/2}\,h_v^{\rm ren}$. From
now on, the superscript ``ren'' will be omitted.

\begin{figure}[htb]
   \vspace{0.5cm}
   \epsfxsize=3.5cm
   \centerline{\epsffile{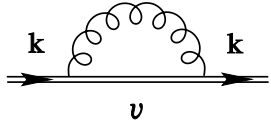}}
\caption{\label{fig:3.4}
One-loop self-energy $-i\Sigma(v\cdot k)$ of a heavy quark in the
HQET.}
\end{figure}

In the minimal subtraction ({\sc ms}) scheme, $Z_h$ can be computed
from the $1/\epsilon$ pole in the heavy-quark self-energy using
\begin{equation}
   1 - Z_h^{-1} = {1\over\epsilon}\mbox{pole of }\,
   {\partial\Sigma(v\cdot k)\over\partial v\cdot k} \,.
\end{equation}
As long as $v\cdot k<0$, the self-energy is infrared finite and real.
The result is gauge-dependent, however. Evaluating the diagram shown
in Fig.~\ref{fig:3.4} in the Feynman gauge, we obtain at one-loop
order
\begin{eqnarray}
   \Sigma(v\cdot k) &=& - i g_s^2\,t_a t_a \int
    {\mbox{d}^D t\over(2\pi)^D}\,{1\over (t^2+i\eta)
     \big[ v\cdot(t+k)+i\eta \big]} \nonumber\\
   &=& - 2i C_F g_s^2 \int\limits_0^\infty\!\mbox{d}\lambda
    \int{\mbox{d}^D t\over(2\pi)^D}\,{1\over\big[ t^2 + 
     2\lambda\,v\cdot(t+k) + i\eta \big]^2} \nonumber\\
   &=& {C_F\alpha_s\over 2\pi}\,\Gamma(\epsilon)
    \int\limits_0^\infty\!\mbox{d}\lambda\,\bigg( 
    {\lambda^2 + \lambda\omega\over 4\pi\mu^2} \bigg)^{-\epsilon} \,,
\end{eqnarray}
where $C_F=4/3$ is a colour factor, $\lambda$ is a dimensionful
Feynman parameter, and $\omega=-2 v\cdot k>0$ acts as an infrared
cutoff. A straightforward calculation leads to
\begin{eqnarray}
   {\partial\Sigma(v\cdot k)\over\partial v\cdot k}
   &=& {C_F\alpha_s\over\pi}\,\Gamma(1+\epsilon)\,
    \bigg( {\omega^2\over 4\pi\mu^2} \bigg)^{-\epsilon}
    \int\limits_0^1\!\mbox{d}z\,z^{-1+2\epsilon}\,
    (1-z)^{-\epsilon} \nonumber\\
   &=& {C_F\alpha_s\over\pi}\,\Gamma(2\epsilon)\,\Gamma(1-\epsilon)\,
    \bigg( {\omega^2\over 4\pi\mu^2} \bigg)^{-\epsilon} \,,
\end{eqnarray} 
where we have substituted $\lambda=\omega\,(1-z)/z$. From an expansion
around $\epsilon=0$, we obtain
\begin{equation}\label{ZfacMS}
   Z_h = 1 + {C_F\alpha_s\over 2\pi\epsilon} \,.
\end{equation}
This result was first derived by Politzer and Wise~\cite{PoWi}. In
the meantime, the calculation was also done at the two-loop
order~\cite{JiMu}$^-$\cite{Gime}.

\subsection{The Residual Mass Term and the Definition of the
Heavy-Quark Mass}

The choice of the expansion parameter in the HQET, i.e.\ the
definition of the heavy-quark mass $m_Q$, deserves some comments. In
the derivation presented earlier in this section, we chose $m_Q$ to
be the ``mass in the Lagrangian'', and using this parameter in the
phase redefinition in (\ref{redef}) we obtained the effective
Lagrangian (\ref{Leff}), in which the heavy-quark mass no
longer appears. However, this treatment has its subtleties. The
symmetries of the HQET allow a ``residual mass term'' $\delta m$ for
the heavy quark, provided that $\delta m$ is of order $\Lambda_{\rm
QCD}$ and is the same for all heavy-quark flavours. Even if we
arrange that such a term is not present at the tree level, it will in
general be induced by quantum corrections. (This is unavoidable if
the theory is regulated with a dimensionful cutoff.) Therefore,
instead of (\ref{Leff}) we should write the effective Lagrangian in
the more general form~\cite{FNL}:
\begin{eqnarray}
   h_v(x) &=& e^{i m_Q v\cdot x}\,P_+\,Q(x) \nonumber\\
   \Rightarrow \qquad
   {\cal L}_\infty &=& \bar h_v\,iv\!\cdot\!D\,h_v
    - \delta m\,\bar h_v h_v \,.
\end{eqnarray}
If we redefine the expansion parameter according to $m_Q\to
m_Q+\Delta m$, the residual mass changes in the opposite way: $\delta
m\to\delta m-\Delta m$. This implies that there is a unique choice of
the expansion parameter such that $\delta m=0$. Requiring $\delta
m=0$, as it is usually done implicitly in the HQET, defines a
heavy-quark mass, which in perturbation theory coincides with the
pole mass~\cite{Tarr}. This, in turn, defines for each heavy hadron a
parameter $\bar\Lambda$ (sometimes called the ``binding energy'')
through
\begin{equation}
   \bar\Lambda = (m_H - m_Q)\Big|_{m_Q\to\infty} \,.
\end{equation}
If one prefers to work with another choice of the expansion
parameter, the values of non-perturbative parameters such as
$\bar\Lambda$ change, but at the same time one has to include the
residual mass term in the HQET Lagrangian. It can be shown that the
various parameters that depend on the definition of $m_Q$ enter the
predictions for all physical observables in such a way that the
results are independent of which particular choice one
adopts~\cite{FNL}.
       
There is one more subtlety hidden in the above discussion. The
quantities $m_Q$, $\bar\Lambda$ and $\delta m$ are non-perturbative
parameters of the HQET, which have a similar status as the vacuum
condensates in QCD phenomenology~\cite{SVZ}. These parameters cannot
be defined unambiguously in perturbation theory. The reason lies in 
the divergent behaviour of perturbative expansions in large orders,
which is associated with the existence of singularities along the
real axis in the Borel plane, the so-called
renormalons~\cite{tHof}$^-$\cite{Muel}. For instance, the
perturbation series which relates the pole mass $m_Q$ of a heavy
quark to its bare mass,
\begin{equation}
   m_Q = m_Q^{\rm bare}\,\Big\{ 1 + c_1\,\alpha_s(m_Q)
   + c_2\,\alpha_s^2(m_Q) + \dots + c_n\,\alpha_s^n(m_Q)
   + \dots \Big\} \,,
\end{equation}
contains numerical coefficients $c_n$ that grow as $n!$ for large
$n$, rendering the series divergent and not Borel
summable~\cite{BBren,Bigiren}. The best one can achieve is to
truncate the perturbation series at the minimal term, but this leads
to an unavoidable arbitrariness of order $\Delta m_Q\sim\Lambda_{\rm
QCD}$ (the size of the minimal term). This observation, which at
first sight seems a serious problem for QCD phenomenology, should
actually not come as a surprise. We know that because of confinement
quarks do not appear as physical states in nature. Hence, there is no
way to define their on-shell properties such as a pole mass. In view
of this, it is actually remarkable that QCD perturbation theory
``knows'' about its incompleteness and indicates, through the
appearance of renormalon singularities, the presence of
non-perturbative effects. We must first specify a scheme how to
truncate the QCD perturbation series before non-perturbative
statements such as $\delta m=0$ become meaningful, and hence before
non-perturbative parameters such as $m_Q$ and $\bar\Lambda$ become
well-defined quantities. The actual values of these parameters will
depend on this scheme.

We stress that the ``renormalon ambiguities'' are not a conceptual
problem for the heavy-quark expansion. In fact, it can be shown quite
generally that these ambiguities cancel in all predictions for
physical observables~\cite{Chris}. The way the cancellations occur is
intricate, however. The generic structure of the heavy-quark
expansion for an observable is of the form:
\begin{equation}
   \mbox{observable} \sim C[\alpha_s(m_Q)]\,\bigg( 1
   + {\Lambda\over m_Q} + \dots \bigg) \,.
\end{equation}
Here $C[\alpha_s(m_Q)]$ represents a perturbative coefficient
function, and $\Lambda$ is a dimensionful non-perturbative parameter.
The truncation of the perturbation series defining the coefficient
function leads to an arbitrariness of order $\Lambda_{\rm QCD}/m_Q$,
which precisely cancels against a corresponding arbitrariness of
order $\Lambda_{\rm QCD}$ in the definition of the non-perturbative
parameter $\Lambda$.

The renormalon problem poses itself when one imagines to apply
perturbation theory in very high orders. In practise, the
perturbative coefficients are known to finite order in $\alpha_s$ (at
best to two-loop accuracy), and to be consistent one should use them
in connection with the pole mass (and $\bar\Lambda$ etc.) defined to
the same order.

\section{Matching and Running}
 
In section~\ref{sec:2}, we have discussed the first two steps in the
construction of the HQET. Integrating out the small components in the
heavy-quark fields, a non-local effective action was derived, which
was then expanded in a series of local operators. The effective
Lagrangian derived that way correctly reproduces the long-distance
physics of the full theory. It does not contain the short-distance
physics correctly, however. The reason is obvious: A heavy quark
participates in strong interactions through its coupling to gluons.
These gluons can be soft or hard, i.e.\ their virtual momenta can be
small, of the order of the confinement scale, or large, of the order
of the heavy-quark mass. But hard gluons can resolve the spin and
flavour quantum numbers of a heavy quark. Their effects lead to a
renormalization of the coefficients of the operators in the HQET. 

Consider, as an example, matrix elements of the vector current
$V=\bar q\,\gamma^\mu Q$. In QCD this current is (partially)
conserved and needs no renormalization~\cite{Prep}. Its matrix
elements are free of ultraviolet divergences. Still, these matrix
elements have a logarithmic dependence on $m_Q$ from the exchange of
hard gluons with virtual momenta of the order of the heavy-quark
mass. If one goes over to the effective theory by taking the limit
$m_Q\to\infty$, these logarithms diverge. Consequently, the vector
current in the effective theory does require a
renormalization~\cite{PoWi}. Its matrix elements depend on an
arbitrary renormalization scale $\mu$, which separates the regions of
short- and long-distance physics. If $\mu$ is chosen such that
$\Lambda_{\rm QCD}\ll\mu\ll m_Q$, the effective coupling constant in
the region between $\mu$ and $m_Q$ is small, and perturbation theory
can be used to compute the short-distance corrections. These
corrections have to be added to the matrix elements of the effective
theory, which contain the long-distance physics below the scale
$\mu$. Schematically, then, the relation between matrix elements in
the full and in the effective theory is
\begin{equation}\label{OPEex}
   \langle V(m_Q)\rangle_{\rm QCD}
   = C_0(m_Q,\mu)\,\langle V_0(\mu)\rangle_{\rm HQET}
   + {C_1(m_Q,\mu)\over m_Q}\,\langle V_1(\mu)\rangle_{\rm HQET}
   + \dots \,,
\end{equation}
where we have indicated that matrix elements in the full theory
depend on $m_Q$, whereas matrix elements in the effective theory are
mass-independent, but do depend on the renormalization scale. The
Wilson coefficients $C_i(m_Q,\mu)$ are defined by this relation.
Order by order in perturbation theory, they can be computed from a
comparison of the matrix elements in the two theories. Since the
effective theory is constructed to reproduce correctly the low-energy
behaviour of the full theory, this ``matching'' procedure is
independent of any long-distance physics, such as infrared
singularities, non-perturbative effects, the nature of the external
states used in the matrix elements, etc.

The calculation of the coefficient functions in perturbation theory
uses the powerful methods of the renormalization group. It is in
principle straightforward, yet in practice rather tedious. A
comprehensive discussion of most of the existing calculations of
short-distance corrections in the HQET can be found in Ref.~8.
Here, we shall discuss as an illustration the renormalization of the
$1/m_Q$-suppressed operators in the effective Lagrangian
(\ref{Lsubl}). At the tree level, there appear two operators at order
$1/m_Q$, which have been given in (\ref{Okin}) and (\ref{Omag}).
Beyond the tree level, the coefficients of these operators may be
modified, and other operators not present at the classical level may
be induced. In general, we thus expect
\begin{equation}
   {\cal L}_{1/m} = C_{\rm kin}(\mu)\,{\cal O}_{\rm kin}(\mu)
   + C_{\rm mag}(\mu)\,{\cal O}_{\rm mag}(\mu)
   + \mbox{new operators} \,,
\label{L5}
\end{equation}
where $\mu$ is the renormalization scale. But how do we calculate the
Wilson coefficient functions, and what are the possible new
operators? To extend the classical construction of
Sec.~\ref{sec:Leff} to include quantum corrections would be
cumbersome. Fortunately, there is a systematic procedure which allows
us to derive the result in the presence of quantum effects in a
rather simple and straightforward way. It consists of three steps:
construction of the operator basis, calculation of the ``matching
conditions'' at $\mu=m_Q$, and renormalization-group improvement
(``running''). Below, we shall first explain these steps in general
and then illustrate them with the particular example of ${\cal
L}_{1/m}$.

\subsection{Construction of the Operator Basis}

Similar to the fields and coupling constants, in a quantum field
theory any composite operator built from quark and gluon fields may
require a renormalization beyond that of its component fields. Such
operators can be divided into three classes: gauge-invariant
operators that do not vanish by the equations of motion (class-I),
gauge-invariant operators that vanish by the equations of motion
(class-II), and operators which are not gauge-invariant (class-III).
In general, operators with the same dimension and quantum numbers mix
under renormalization. However, things simplify if one works with the
background field technique~\cite{DeWi}$^-$\cite{Abbo}, which is an
elegant method for quantizing gauge theories, preserving explicit
gauge invariance. This offers the advantage that a class-I operator
cannot mix with class-III operators, so that only gauge-invariant
operators need to be considered~\cite{Klug}. Furthermore, class-II
operators are irrelevant since their matrix elements vanish by the
equations of motion. It it thus sufficient to consider class-I
operators only.

Thus, we must find a complete set of class-I operators of the right
dimension, carrying the quantum numbers allowed by the symmetries of
the problem. In the case at hand, we are dealing with operators
appearing at order $1/m_Q$ in a strong-interaction Lagrangian, and we
thus have to find dimension-five operators containing two heavy-quark
fields of the same velocity. Moreover, these operators must transform
as scalars under the Lorentz group. The most general form of such
operators is
\begin{equation}
   \bar h_v\,\Gamma_{\mu\nu}\,iD^\mu iD^\nu\,h_v \,; \qquad
   \Gamma_{\mu\nu} \in \Big\{ g_{\mu\nu}, v_\mu v_\nu,
   \gamma_\mu v_\nu, \gamma_\nu v_\mu,
   \textstyle\frac 12\,[\gamma_\mu,\gamma_\nu] \Big\} \,.
\end{equation}
Note that the velocity is not a dynamical quantity in the HQET and
thus can be used to construct the basis operators. Using that $\bar
h_v\gamma_\mu\,h_v = \bar h_v\,v_\mu\,h_v$, we find that there are
only three possible operators:
\begin{equation}
   \bar h_v\,(iD)^2 h_v \,, \qquad
   \bar h_v\,(iv\!\cdot\!D)^2 h_v \,, \qquad
   \textstyle\frac 12\,\bar h_v\,\sigma_{\mu\nu} g_s G^{\mu\nu} h_v
   \,.
\end{equation}
Since the equation of motion of the HQET is $iv\!\cdot\!D\,h_v=0$, it
follows that there are two class-I and one class-II operators, which
we choose in the form:
\begin{eqnarray}
   \mbox{class-I:} \quad
   &&\bar h_v\,(iD_\perp)^2 h_v \,, \qquad
    \textstyle\frac 12\,\bar h_v\,\sigma_{\mu\nu} g_s G^{\mu\nu} h_v
    \,, \nonumber\\
   \mbox{class-II:} \quad
   &&\bar h_v\,(iv\!\cdot\!D)^2 h_v \,.
\end{eqnarray}
Besides a class-II operator, which has vanishing matrix elements
between physical states, the kinetic and chromo-magnetic operators
already present at the tree level are thus the only operators which
can appear in ${\cal L}_{1/m}$, even in the presence of quantum
corrections. Once we have found a complete basis of class-I
operators, our next goal is to calculate their coefficient functions
in perturbation theory.

\subsection{Matching Conditions at $\mu=m_Q$}

The Wilson coefficient functions $C_{\rm kin}(\mu)$ and $C_{\rm
mag}(\mu)$ in (\ref{L5}) can be obtained from the comparison
(``matching'') of Green's functions in QCD with those in the
effective theory. It is crucial that, by construction, the Wilson
coefficients receive only short-distance contributions (see
Fig.~\ref{fig:magic}) and are thus insensitive to the properties of
the external states. This ensures that once the coefficients have
been determined by requiring that some particular Green's function(s)
be the same in the two theories, all other Green's functions will be
the same. Moreover, since the Wilson coefficients are infrared
insensitive they are calculable in perturbation theory, and we can
perform their calculation using quark and gluon states rather than
physical hadron states.

\begin{figure}[htb]
   \epsfxsize=8cm
   \centerline{\epsffile{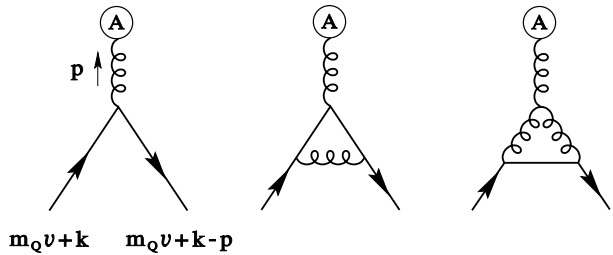}}
   \centerline{\parbox{11cm}{\caption{\label{fig:glue}
Diagrams for the calculation of the heavy quark-gluon vertex function
in QCD. The background field is denoted by $A$.
   }}}
\end{figure}

In the example at hand, the coefficients $C_{\rm kin}(\mu)$ and
$C_{\rm mag}(\mu)$ can be obtained from a calculation of the Green's
function of two heavy quarks and a background gluon field, to
one-loop order in the full and in the effective theory. The relevant
vertex diagrams in QCD are shown in Fig.~\ref{fig:glue}. They have to
be supplemented by the wave-function renormalization of the external
quark lines. The background field is not renormalized. The momentum
assignments are such that $p$ is the outgoing momentum of the
background field, and $k$ and $(k-p)$ are the residual momenta of the
heavy quarks. To order $1/m_Q$, it is sufficient to keep terms linear
in $k$ or $p$. The quarks can be taken on shell, in which case
$v\cdot k=v\cdot p=0$. A subtlety which has to be taken into account
is that the QCD spinor $u_Q(P_Q,s)$ is related to the spinor
$u_h(v,s)$ of the effective theory by
\begin{equation}
   u_Q(P_Q,s) = \bigg( 1 + {\rlap/k\over 2 m_Q} + \ldots \bigg)\,
   u_h(v,s) \,,
\end{equation}
where $P_Q = m_Q v + k$. In the matching calculation one has to use 
the same spinors in both theories. We thus define a vertex function
$\Gamma^\mu$ by writing the amplitude as $i\mu^\epsilon g_s
A_{\mu,a}(p)\,\bar u_h \Gamma^\mu t_a u_h$, so that at the tree level
in QCD
\begin{eqnarray}\label{GQCD0}
   \Gamma_{{\rm QCD},0}^\mu &=& \bigg( 1 
    + {\rlap/k - \,\rlap/\!p\over 2 m_Q} \bigg)\,\gamma^\mu\,
    \bigg( 1 + {\rlap/k\over 2 m_Q} \bigg) + \ldots \nonumber\\
   &=& v^\mu + {(2 k-p)^\mu\over 2 m_Q} 
    + {\big[ \gamma^\mu,\,\rlap/\!p \big]\over 4 m_Q} + \ldots \,.
\end{eqnarray}
Here the ellipses represent terms of higher order in $k$ or $p$, and
we have used that between the heavy-quark spinors $\gamma^\mu$ can be
replaced by $v^\mu$.

The contributions to the vertex function arising at the one-loop
level are also shown in Fig.~\ref{fig:glue}. They contain both
abelian and non-abelian vertices. Since the matching calculation is
insensitive to any long-distance properties such as the nature of the
infrared regulator, it is legitimate to work with any infrared
regularization scheme that is convenient. Following Eichten and
Hill~\cite{EiH2,MN94}, we choose to regulate both ultraviolet and
infrared divergences using dimensional regularization. Moreover, we
expand the resulting expressions for the Feynman amplitudes to linear
order in the external momenta and then set the external momenta to
zero inside the loop integrals. Then the only mass scale remaining is
the heavy-quark mass. In the $\overline{\mbox{\sc ms}}$ scheme, the
result for the one-loop contribution to the QCD vertex function
is~\cite{EiH2}
\begin{equation}
   \Gamma_{{\rm QCD},1}^\mu 
   = {\big[ \gamma^\mu,\,\rlap/\!p \big]\over 4 m_Q}\,
   {\alpha_s\over 2\pi}\,\bigg( - C_A \ln{m_Q\over\mu} + C_A + C_F
   \bigg) \,,
\end{equation}
where $C_F=\frac 12(N_c^2-1)/N_c=4/3$ and $C_A=N_c=3$ are the
eigenvalues of the quadratic Casimir operator in the fundamental and
the adjoint representations.

Now comes a clue: if dimensional regularization is used to regulate
both ultraviolet and infrared singularities, all loop integrals in
the HQET are no-scale integrals (after a power of the external
momenta has been factored out) and vanish! So only the tree-level
matrix elements of the HQET operators in the effective Lagrangian
multiplied by their Wilson coefficient functions remain. This is why
dimensional regularization is superb for matching calculations. The
result is
\begin{equation}\label{GHQET}
   \Gamma_{\rm HQET}^\mu = v^\mu
   + C_{\rm kin}(\mu)\,{(2 k-p)^\mu\over 2 m_Q} 
   + C_{\rm mag}(\mu)\,{\big[ \gamma^\mu,\,\rlap/\!p \big]
   \over 4 m_Q} + \dots \,.
\end{equation}
Requiring that the vertex functions be the same in the full and in the
effective theory, we find (in the $\overline{\mbox{\sc ms}}$ scheme)
\begin{equation}
   C_{\rm kin}(\mu) = 1 \,, \qquad
   C_{\rm mag}(\mu) = 1 + {\alpha_s\over 2\pi}\, 
   \bigg( - C_A \ln{m_Q\over\mu} + C_A + C_F \bigg) \,.
\label{CConel}
\end{equation}
The fact that the kinetic operator is not renormalized is not an
accident, but follows from an invariance of the HQET under small
redefinitions of the velocity used in the construction of the
effective Lagrangian. Clearly, the predictions of the HQET should not
depend on whether $v$ is taken to be the velocity of the hadron
containing the heavy quark, the velocity of the heavy quark itself,
or some other velocity differing from the hadron velocity by an
amount of order $\Lambda_{\rm QCD}/m_Q$. This so-called
reparametrization invariance implies that $C_{\rm kin}(\mu)=1$ must
hold to all orders in perturbation theory~\cite{luma,cino}.

In the next paragraph, we will see that the scale dependence
predicted by the one-loop result quoted above cannot be trusted if
$\mu\ll m_Q$; however, what can be obtained from the matching
calculation are the values of the coefficient functions at the
matching scale $\mu=m_Q$ as well as their logarithmic derivatives.
For the coefficient of the chromo-magnetic operator, we find:
\begin{eqnarray}
   C_{\rm mag}(m_Q) &=& 1 + (C_A + C_F)\,{\alpha_s(m_Q)\over 2\pi}
    \,, \nonumber\\
   {\mbox{d}\ln C_{\rm mag}(\mu)\over\mbox{d}\ln\mu}
   &=& C_A\,{\alpha_s\over 2\pi} \,.
\label{Cmmat}
\end{eqnarray}

\subsection{Renormalization-Group Evolution}

The one-loop calculation presented above allows us to derive
expressions for the Wilson coefficient functions provided that
$m_Q/\mu=O(1)$. In practical applications of effective field
theories, one is however often interested in the case where there is a
large ratio of mass scales. After all, an effective theory is
constructed to separate the physics on two very different energy
scales. In such a situation, the coefficient functions contain large
logarithms of the type $[\alpha_s\ln(m_Q/\mu)]^n$, which must be
summed to all orders in perturbation theory. This is achieved by
using the powerful machinery of the renormalization group.

For a set $\{{\cal O}_i\}$ of $n$ class-I operators that mix under
renormalization, one defines an $n\times n$ matrix of renormalization
factors $Z_{ij}$ by ${\cal O}_i^{\rm bare} = Z_{ij}\,{\cal
O}_j(\mu)$, such that the matrix elements of the renormalized
operators ${\cal O}_j(\mu)$ remain finite as $\epsilon\to 0$. In
contrast to the bare operators, the renormalized ones depend on the
subtraction scale via the $\mu$ dependence of $Z_{ij}$:
\begin{equation}
   \mu {{\rm d}\over{\rm d}\mu}\,{\cal O}_i(\mu)
   = \bigg(\mu {{\rm d}\over{\rm d}\mu} Z_{ij}^{-1}\bigg)\,
   {\cal O}_j^{\rm bare} = - \gamma_{ik}\,{\cal O}_k(\mu) \,,
\end{equation}
where
\begin{equation}
   \gamma_{ik} = - \bigg(\mu {{\rm d}\over{\rm d}\mu}
   Z_{ij}^{-1}\bigg)\,Z_{jk}
   = Z_{ij}^{-1}\,\mu {{\rm d}\over{\rm d}\mu}\,Z_{jk}
\label{anom}
\end{equation}
are called the anomalous dimensions. That under a change of the
renormalization scale the operators mix among themselves follows from
the fact that the basis of operators is complete. It is convenient to
introduce a compact matrix notation, in which $\vec{\cal O}(\mu)$ is
the vector of renormalized operators, $\hat Z$ is the matrix of
renormalization factors, and $\hat\gamma$ denotes the anomalous
dimension matrix. Then the scale dependence of the renormalized
operators is controlled by the renormalization-group equation (RGE)
\begin{equation}\label{RGEops}
   \bigg( \mu {{\rm d}\over{\rm d}\mu} + \hat\gamma \bigg)\,
   \vec{\cal O}(\mu) = 0 \,.
\end{equation}
In the $\mbox{\sc ms}$ scheme, the matrix $\hat Z$ obeys an expansion
of the form
\begin{equation}
   \hat Z = 1 + \sum_{k=1}^\infty {1\over\epsilon^k}\,
   \hat Z_k(\alpha_s) \,,
\end{equation}
and by requiring that the anomalous dimensions in (\ref{anom}) be
finite as $\epsilon\to 0$ one finds that $\hat\gamma$ can be computed
in terms of the coefficient of the $1/\epsilon$ pole~\cite{DGro}:
\begin{equation}\label{gamZ1}
   \hat\gamma = - 2\alpha_s\,
   {\partial\hat Z_1(\alpha_s)\over\partial\alpha_s} \,.
\end{equation}
The same relation holds in the $\overline{\mbox{\sc ms}}$ scheme.

From (\ref{RGEops}) and the fact that the product $C_i(\mu)\,{\cal
O}_i(\mu)$ must be $\mu$ independent, we derive the RGE satisfied by
the coefficient functions. It reads
\begin{equation}\label{RGE}
   \bigg(\mu {{\rm d}\over{\rm d}\mu} - \hat\gamma^T \bigg)\,
   \vec C(\mu) = 0 \,,
\end{equation}
where we have collected the coefficients into a vector $\vec C(\mu)$.
In general, the Wilson coefficients can depend on $\mu$ both
explicitly or implicitly through the running coupling. We thus have
\begin{equation}
   \mu {{\rm d}\over{\rm d}\mu} = \mu {\partial\over\partial\mu}
   + \beta(\alpha_s)\,{\partial\over\partial \alpha_s(\mu)} \,,
\end{equation}
where the $\beta$ function
\begin{equation}
   \beta\big(\alpha_s) = \mu\,
   {\partial\alpha_s(\mu)\over\partial\mu}
   = -2\alpha_s\,\bigg[\, \beta_0\,{\alpha_s\over 4\pi}
   + \beta_1\,\bigg( {\alpha_s\over 4\pi} \bigg)^2 + \ldots \bigg]
\label{betaf}
\end{equation}
describes the scale dependence of the renormalized coupling constant.
The one- and two-loop coefficients are scheme independent and are
given by~\cite{Gros,Poli,Bela}
\begin{eqnarray}
   \beta_0 &=& {11\over 3}\,C_A - {4\over 3}\,T_F\,n_f \,,
    \nonumber\\
   \beta_1 &=& {34\over 3}\,C_A^2 - \left( {20\over 3}\,C_A + 4 C_F
    \right) T_F\,n_f \,,
\end{eqnarray}
where $n_f$ is the number of light quark flavours, and $T_F=1/2$ is
the normalization of the SU(3) generators in the fundamental
representation: $\mbox{tr}(t_a t_b)=T_F\,\delta_{ab}$. It is now
straightforward to obtain a formal solution of the RGE. It reads
\begin{equation}\label{RGEsol}
   \vec C(\mu) = \hat U(\mu,m_Q)\,\vec C(m_Q) \,, 
\end{equation}
with the evolution matrix~\cite{Bura}$^-$\cite{BJLW}
\begin{equation}\label{Uevol}
   \hat U(\mu,m_Q) = T_\alpha\,\exp\!
   \int\limits_{\displaystyle\alpha_s(m_Q)}
    ^{\displaystyle\alpha_s(\mu)}\!
   {\rm d}\alpha\,{\hat\gamma^T(\alpha)\over\beta(\alpha)} \,.
\end{equation}
Here ``$T_\alpha$'' means an ordering in the coupling constant such
that the couplings increase from right to left (for $\mu<m_Q$). This
is necessary since, in general, the anomalous dimension matrices at
different values of $\alpha_s$ do not commute. Eq.~(\ref{Uevol}) can
be solved perturbatively by expanding the $\beta$ function
(\ref{betaf}) and the anomalous dimension matrix in powers of the
renormalized coupling constant:
\begin{equation}
   \hat\gamma(\alpha_s) = \hat\gamma_0\,{\alpha_s\over 4\pi} 
   + \hat\gamma_1\,\bigg( {\alpha_s\over 4\pi} \bigg)^2
   + \dots \,. 
\end{equation}
Here we shall only discuss the important case of a single coefficient
function, or equivalently, when there is no operator mixing. Then the
matrix $\hat\gamma$ reduces to a number, and the evolution is
described by a function $U(\mu,m_Q)$, for which the perturbative
solution of (\ref{Uevol}) at next-to-leading order yields
\begin{equation}\label{UNLO}
   U_{\rm NLO}(\mu,m_Q) = \left( {\alpha_s(m_Q)\over\alpha_s(\mu)}
   \right)^a \left\{ 1 + {\alpha_s(m_Q)-\alpha_s(\mu)\over 4\pi}\,S
   + \dots \right\} \,,
\end{equation}
with
\begin{equation}\label{aSNLO}
   a = {\gamma_0\over 2\beta_0} \,, \qquad
   S = {\gamma_1\over 2\beta_0}
   - {\gamma_0\beta_1\over 2\beta_0^2} \,.
\end{equation}

The theoretical framework discussed here is called
``renormalization-group (RG) improved perturbation theory''. In the
expression for the evolution function $U(\mu,m_Q)$, there are no
large logarithms of the form $\alpha_s\ln(m_Q/\mu)$ left. They are
all contained in the ratio of the running couplings evaluated at the
scales $m_Q$ and $\mu$. Thus, RG-improved perturbation theory
provides the optimal method to bridge wide energy intervals. (As a
side remark, we note that the same technique is used to control the
evolution of gauge couplings and running mass parameters from low
energies up to very high energy scales characteristic of grand
unified theories.) The terms shown explicitly in (\ref{UNLO})
correspond to the so-called next-to-leading order (NLO) in
RG-improved perturbation theory. In this approximation, the leading
and subleading logarithms $[\alpha_s\ln(m_Q/\mu)]^n$ and $\alpha_s
[\alpha_s\ln(m_Q/\mu)]^n$ are summed correctly to all orders in
perturbation theory. To achieve this, it is necessary to calculate
the two-loop coefficient $\gamma_1$ of the anomalous dimension. When
$\gamma_1$ is not known, it is only possible to evaluate the
evolution function in the so-called leading logarithmic order (LO),
in which
\begin{equation}
   U_{\rm LO}(\mu,m_Q) = \left(
   {\alpha_s(m_Q)\over\alpha_s(\mu)} \right)^a \,.
\end{equation}
This still sums the leading logarithms to all orders, but does not
contain the non-logarithmic terms of order $\alpha_s$. 

To complete the calculation of the RG-improved coefficient function,
the evolution function $U(\mu,m_Q)$ must be combined with the initial
condition for the Wilson coefficient at the high energy scale
$\mu=m_Q$, as shown in (\ref{RGEsol}). If the operator under
consideration is present at the tree level, the matching condition
can be written in the form
\begin{equation}
   C(m_Q) = 1 + c_1\,\frac{\alpha_s(m_Q)}{4\pi} + \dots \,,
\end{equation}
where $c_1$ is obtained from a one-loop calculation. To obtain a
consistent (i.e.\ renormalization-scheme independent) result at
next-to-leading order, we have to combine the one-loop matching
condition with the expression for $U(\mu,m_Q)$ given in (\ref{UNLO}).
This requires the calculation of the two-loop anomalous dimension.
The result is
\begin{equation}\label{Csolu}
   C_{\rm NLO}(\mu) = \left( {\alpha_s(m_Q)\over\alpha_s(\mu)}
   \right)^a \left\{ 1 + {\alpha_s(m_Q)\over 4\pi}\,(S+c_1)
   - {\alpha_s(\mu)\over 4\pi}\,S \right\} \,.   
\end{equation}
In this expression, the terms involving the coupling constant
$\alpha_s(m_Q)$ are renormalization-scheme
independent~\cite{Bura,Flor}. The exponent $a$ involves only the
one-loop coefficients $\gamma_0$ and $\beta_0$ and is scheme
independent by itself. For the coefficient $(S+c_1)$ of the
next-to-leading term things are more complicated, however. The
one-loop matching coefficient $c_1$, the two-loop anomalous dimension
$\gamma_1$, and the QCD scale parameter $\Lambda_{\rm QCD}$ in the
expression for the running coupling constant are all scheme
dependent, but they conspire to give $\alpha_s(m_Q)$ a
scheme-independent coefficient. On the other hand, the coefficient
$S$ of $\alpha_s(\mu)$ does depend on the renormalization procedure.
This is not a surprise; only when the $\mu$-dependent terms in the
Wilson coefficient are combined with the $\mu$-dependent matrix
elements of the renormalized operator one can expect to obtain a
scheme-independent result. For this reason, it is sometimes useful to
factorize the solution (\ref{Csolu}) in the form $C(\mu)\equiv
\widehat{C}(m_Q)\,K(\mu)$, where $\widehat{C}(m_Q)$ is RG-invariant
and contains all dependence on the large mass scale $m_Q$. The
scheme-dependent function $K(\mu)$ can be used to define a
RG-invariant renormalized operator: $\widehat{\cal O}\equiv
K(\mu)\,{\cal O}(\mu)$. At next-to-leading order, we obtain:
 \begin{eqnarray}
   \widehat{C}(m_Q) &=& \left[ \alpha_s(m_Q) \right]^a\,
    \left\{ 1 + {\alpha_s(m_Q)\over 4\pi}\,(S+c_1) \right\} \,,
    \nonumber\\
   \widehat{\cal O} &=& \left[ \alpha_s(\mu) \right]^{-a}\,
    \left\{ 1 - {\alpha_s(\mu)\over 4\pi}\,S \right\}\,{\cal O}(\mu)
    \,.
\end{eqnarray}    

Let us finally apply this formalism to the operators appearing at
order $1/m_Q$ in the effective Lagrangian of the HQET. The fact that
reparametrization invariance ensures that the kinetic operator is not
renormalized implies that the $2\times 2$ anomalous dimension matrix
for the operators ${\cal O}_{\rm kin}$ and ${\cal O}_{\rm mag}$ is
diagonal and of the form
\begin{equation}
   \hat\gamma = \left( \begin{array}{cc}
   0 & 0 \\ 0 & \gamma^{\rm mag} \end{array} \right) \,.
\end{equation}
The one-loop coefficient of the anomalous dimension of the
chromo-magnetic operator, together with the one-loop matching
coefficient, can be obtained from (\ref{Cmmat}):
\begin{equation}
   \gamma_0^{\rm mag} = 2 C_A \,,\qquad 
   c_1^{\rm mag} = 2(C_A + C_F) \,.
\end{equation}       
The result for $\gamma_0^{\rm mag}$ can also be obtained in a simpler
way by computing only the $1/\epsilon$ poles in the matrix elements
of the bare operators~\cite{FGL}. Unfortunately, the two-loop
coefficient $\gamma_1^{\rm mag}$ is not yet known.\footnote{This is
one of the few cases of interest where an anomalous dimension in the
HQET is not yet known to two-loop order. If you feel strong enough,
you are invited to try the calculation of $\gamma_1^{\rm mag}$!} This
means that the coefficient $S_{\rm mag}$ in the next-to-leading order
solution (\ref{Csolu}) is still unknown.

\subsection{Renormalization of Heavy-Quark Currents}

As a final example, we discuss the renormalization of local current
operators involving two heavy-quark fields. This case is of
particular importance for phenomenology, as the weak current for
$b\to c\,\ell\,\bar\nu$ transitions is of this form. In the HQET, the
relevant current operators contain two heavy-quark fields at
different velocity and are thus of the form $\bar h_{v'}\Gamma\,h_v$
(it does not matter whether the two fields have the same flavour),
where $\Gamma$ is some Dirac matrix, whose structure is irrelevant to
our discussion. We have discussed in Sec.~\ref{sec:3} that the matrix
elements of such operators between meson states are proportional to
the universal Isgur--Wise form factor $\xi(v\cdot v')$. We shall now
derive, in leading logarithmic order, the Wilson coefficient function
that relates the QCD current operators with their HQET counterparts
renormalized at the scale $\mu\ll m_Q$.

\begin{figure}[htb]
   \epsfxsize=3cm
   \centerline{\epsffile{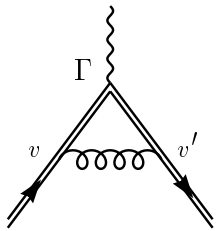}}
   \centerline{\parbox{11cm}{\caption{\label{fig:hh}
One-loop vertex diagram arising in the calculation of the anomalous
dimension of heavy-quark currents. The external current changes the
heavy-quark velocity from $v$ to $v'$.
   }}}
\end{figure}

To this end, we need to calculate, using dimensional regularization,
the $1/\epsilon$ pole in the matrix element of the bare current
operator between quark states. The relevant vertex diagram is shown
in Fig.~\ref{fig:hh}. Since in the effective theory the coupling of a
heavy quark to a gluon does not involve a $\gamma$ matrix, it is easy
to see that to all orders in perturbation theory the operator $\bar
h_{v'}\Gamma\,h_v$ is renormalized multiplicatively and irrespective
of its Dirac structure. The extraction of the one-loop ultraviolet
divergence can be done in a few lines. In the Feynman
gauge,\footnote{The final result for the anomalous dimension is gauge
independent.}
the value of the vertex diagram is (omitting the quark spinors):
\begin{eqnarray}
   && - 4i g_s^2 t_a t_a\,v\cdot v'\,\Gamma \int
    {\mbox{d}^D t\over(2\pi)^D}\,{1\over (t^2+i\eta)
     (v\cdot t+i\eta) (v'\cdot t+i\eta)} \nonumber\\
   &=& - 4i g_s^2 C_F\,v\cdot v'\,\Gamma
    \int\limits_0^\infty\!\mbox{d}\lambda
    \int\limits_0^\infty\!\mbox{d}\rho
    \int{\mbox{d}^D t\over(2\pi)^D}\,{1\over\big[ t^2 + 
     2(\rho v+\lambda v')\cdot t + i\eta \big]^3} \nonumber\\
   &=& - {C_F\alpha_s\over\pi}\,\Gamma(1+\epsilon)\,
    v\cdot v'\,\Gamma\,(4\pi\mu^2)^\epsilon
    \int\limits_0^\infty\!\mbox{d}\lambda
    \int\limits_0^\infty\!\mbox{d}\rho\,
    {1\over \left(\rho^2 + \lambda^2
     + 2 v\cdot v'\,\rho\lambda \right)^{1+\epsilon}} \,.
    \nonumber\\
\end{eqnarray}
Defining a new variable $z=\rho/\lambda$, and introducing an
arbitrary infrared cutoff $\delta$, we can rewrite the double
integral in the form:
\begin{equation}
    \int\limits_0^\infty\!\mbox{d}\lambda\,
    {\lambda\over \left(\lambda^2 + \delta^2 \right)^{1+\epsilon}}
    \int\limits_0^\infty\!\mbox{d}z\,
    {1\over \left(1 + z^2 + 2 w z \right)^{1+\epsilon}}
    = {\delta^{-2\epsilon}\over 2\epsilon}\,r(w)
    + \mbox{finite~terms,}
\end{equation}
where $w=v\cdot v'$, and
\begin{equation}
   r(w) = \int\limits_0^\infty\!\mbox{d}z\,{1\over 1 + z^2 + 2 w z}
   = {1\over\sqrt{w^2-1}}\,\ln\left(w + \sqrt{w^2-1}\right) \,.
\end{equation}
Hence, the $1/\epsilon$ pole of the vertex diagram is given by
\begin{equation}
    - {C_F\alpha_s\over 2\pi\epsilon}\,\Gamma\,w\,r(w) \,.
\end{equation}
To obtain the renormalization constant $Z_{hh}$ of the bare current
operator, we have to add a contribution
\begin{equation}
    Z_h\,\Gamma = \left( 1 + {C_F\alpha_s\over 2\pi\epsilon}
    \right) \Gamma 
\end{equation}
from the wave-function renormalization of the heavy-quark fields,
where $Z_h$ has been given in (\ref{ZfacMS}). The result is
\begin{equation}
   Z_{hh} = 1 - {C_F\alpha_s\over 2\pi\epsilon} 
   \left[ w\,r(w)-1 \right] + \mbox{finite~terms.}
\end{equation}
By means of the relation (\ref{gamZ1}), we derive from this the
one-loop coefficient of the anomalous dimension of heavy-quark
currents in the HQET. This is the famous velocity-dependent anomalous
dimension obtained by Falk et al.~\cite{Falk}:
\begin{equation}\label{gam0hh}
   \gamma_0^{\rm hh}(w) = 4 C_F \left[ w\,r(w)-1 \right] \,,
\end{equation}
In the zero-recoil limit, i.e.\ for $w=1$, the heavy-quark currents
are the symmetry currents of the spin--flavour symmetry, and as such
they are not renormalized, since the associated charges are
conserved. This implies that
\begin{equation}
   \gamma^{\rm hh}(1) = 0
\end{equation}    
to all orders in perturbation theory. This constraint is satisfied by
the one-loop result in (\ref{gam0hh}), since $r(1)=1$.

Since in the effective theory the velocity of a heavy quark is
conserved by the strong interactions, the heavy quark can be
described by a Wilson line~\cite{EiFe}. An external current can
instantaneously change the velocity, resulting in a kink of that
line. It is well-known that such cusps lead to singular behaviour.
The renormalization of cusp singularities of Wilson lines was
investigated in detail by Korchemsky and Radyushkin~\cite{KoRa}
already in 1987, prior to the development of the HQET. In particular,
they calculated the one- and two-loop coefficients of the so-called
cusp anomalous dimension $\gamma^{\rm cusp}(\varphi)$ as a function
of the hyperbolic cusp angle $\varphi$. But this anomalous dimension
is precisely that of heavy-quark currents~\cite{Korc}, with the
identification $\cosh\varphi=w$. Later, the result for $\gamma_1^{\rm
hh}(w)$, which we will not present here, has been confirmed in the
context of the HQET~\cite{Kili}.

\begin{figure}[htb]
   \epsfxsize=6.5cm
   \centerline{\epsffile{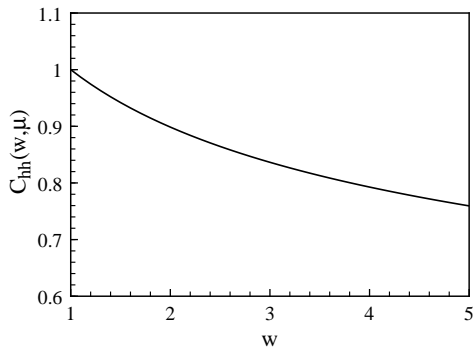}}
   \centerline{\parbox{11cm}{\caption{\label{fig:Chh}
Velocity dependence of the Wilson coefficient $C_{\rm hh}(w,\mu)$,
evaluated for $\alpha_s(\mu)/\alpha_s(m_Q)=2$ and $n_f=3$.
   }}}
\end{figure}

In leading logarithmic order, the expansion of heavy-quark currents
in the HQET takes the form
\begin{equation}
   \bar Q\,\Gamma\,Q \to C_{\rm hh}(w,\mu)\,\bar h_{v'}\Gamma\,h_v
   + O(1/m_Q) \,,
\end{equation}
where
\begin{equation}
   C_{\rm hh}(w,\mu) = \left( {\alpha_s(m_Q)\over\alpha_s(\mu)}
   \right)^{a_{\rm hh}(w)} \,,\qquad
   a_{\rm hh}(w) = {2 C_F\over\beta_0}\,\left[ w\,r(w) - 1
   \right] \,.
\label{Chh}
\end{equation}
The velocity dependence of the Wilson coefficient is illustrated in Fig.~\ref{fig:Chh}. Physical decay or scattering amplitudes are
proportional to the product
\begin{equation}
   C_{\rm hh}(w,\mu)\,\xi(w,\mu) \,,
\end{equation}
where $\xi(w,\mu)$ is the renormalized Isgur--Wise function, which
satisfies $\xi(1,\mu)=1$. The fact that $a_{\rm hh}(1)=0$ implies
that RG effects respect the normalization of form factors at zero
recoil.

We like to finish this discussion with a curious remark, which
illustrates the power of RG methods. For values $w=O(1)$, the
leading-order expression for the coefficient $C_{\rm hh}(w,\mu)$ in
(\ref{Chh}) contains all large logarithms of the type
$[\alpha_s\ln(m_Q/\mu)]^n$. For very large values of $w$, however, we
have $w\,r(w)\to\ln(2w)-1$, so that in the RG improvement of $C_{\rm
hh}(w,\mu)$ we have resummed terms of the form $[\alpha_s\ln
w\ln(m_Q/\mu)]^n$. These are the well-known Sudakov double
logarithms, which arise from the emission of gluon bremsstrahlungs
during the scattering of the heavy quarks. This effect leads to a
fractional power-like damping of the transition form factors at large
recoil, which adds to the ``soft'' suppression contained in the
Isgur--Wise form factor itself~\cite{Groz}. Explicitly, we obtain
\begin{equation}
   C_{\rm hh}(w,\mu) \to \left( \frac{e}{2 w} \right)^\eta \,,\qquad
   \eta = \frac{2 C_F}{\beta_0}
   \ln\frac{\alpha_s(\mu)}{\alpha_s(m_Q)} \,.
\end{equation}

\section{Concluding Remarks}
 
We have presented an introduction to heavy-quark symmetry, the
heavy-quark effective theory and the $1/m_Q$ expansion, which
provide the modern theoretical tools to perform quantitative
calculations in heavy-flavour physics. Our hope was to convince the
reader that heavy-flavour physics is a rich and diverse area of
research, which is at present characterized by a fruitful interplay
between theory and experiments. This has led to many significant
discoveries and developments on both sides. Heavy-quark physics has
the potential to determine many important parameters of the
electroweak theory and to test the Standard Model at low energies. At
the same time, it provides an ideal laboratory to study the nature of
non-perturbative phenomena in QCD, still one of the least understood
properties of the Standard Model.

Let us finish with a somewhat philosophical remark: At this school,
we have heard a lot about exciting new developments related to
dualities, which relate apparently very different theories to each
other. So are electric, weak-coupling phenomena in one theory dual to
magnetic, strong-coupling phenomena in another theory. Some people
argue quite convincingly that duality seems to be everywhere in
nature, and consequently there are no really difficult questions in
physics; very difficult problems become trivial when approached from
a different, dual point of view. There are, however, ``moderately
difficult'' problems in physics, which are ``self-dual''. It is the
author's opinion that real-world (i.e.\ non-supersymmetric) QCD at
hadronic energies belongs to this category. Having said this, we
conclude that heavy-quark effective theory provides a powerful tool
to tackle the ``moderately difficult'' problems of heavy-flavour
physics.

\section*{Acknowledgments}
It is my pleasure to thank the organizers and the staff of the
International School of Subnuclear Physics
for the invitation to present these lectures and for 
making my stay in Erice
such an enjoyable one.

\end{document}